\documentclass[prd,preprint,tightenlines,showpacs,preprintnumbers,nofootinbib,eqsecnum,superscriptaddress]{revtex4-1}

 \usepackage[dvips,final]{graphicx}
  \usepackage{amssymb}
   \usepackage{amsmath}
    \usepackage{amsfonts}
     \usepackage{epsfig}
      \usepackage{bm}

\usepackage[section]{placeins}

\usepackage{sidecap}
\usepackage{multirow}
\usepackage{ctable}
\usepackage{booktabs}
\usepackage{array}
\usepackage{tabularx}
\usepackage{xcolor}
\usepackage{pstricks}

\newcommand{\bl}{\mbox{\boldmath $l$}}
\newcommand{\bp}{\mbox{\boldmath $p$}}
\newcommand{\ba}{\mbox{\boldmath $a$}}
\newcommand{\bq}{\mbox{\boldmath $q$}}
\newcommand{\be}{\mbox{\boldmath $e$}}
\newcommand{\bP}{\mbox{\boldmath $P$}}

\newcommand{\bk}{\mbox{\boldmath $k$}}

\newcommand{\Tr}{{\mbox{\rm Tr}}}

\newcommand{\ket}[1]{ {#1} \rangle}
\newcommand{\bra}[1]{\langle {#1} }

\newcommand{\half}{{1\over 2}}

\begin{document}


\title{Hadroproduction of 
scalar $P$-wave quarkonia\\ 
in the light-front $k_T$-factorization approach
}

\author{Izabela Babiarz}
\email{izabela.babiarz@ifj.edu.pl.pl}
\affiliation{Institute of Nuclear Physics, Polish Academy of Sciences, 
ul. Radzikowskiego 152, PL-31-342 Krak{\'o}w, Poland}

\author{Roman Pasechnik}
\email{roman.pasechnik@thep.lu.se}
\affiliation{Department of Astronomy and Theoretical Physics,
Lund University, SE-223 62 Lund, Sweden}
\affiliation{Nuclear Physics Institute ASCR, 25068 Re\v{z}, Czech Republic}
\affiliation{Departamento de F\'isica, CFM, Universidade Federal 
de Santa Catarina, C.P. 476, CEP 88.040-900, Florian\'opolis, 
SC, Brazil}

\author{Wolfgang Sch\"afer}
\email{wolfgang.schafer@ifj.edu.pl} 
\affiliation{Institute of Nuclear
Physics, Polish Academy of Sciences, ul. Radzikowskiego 152, PL-31-342 
Krak{\'o}w, Poland}

\author{Antoni Szczurek}
\email{antoni.szczurek@ifj.edu.pl}
\affiliation{Institute of Nuclear
Physics, Polish Academy of Sciences, ul. Radzikowskiego 152, PL-31-342 
Krak{\'o}w, Poland}
\affiliation{Faculty of Mathematics and Natural Sciences,
University of Rzesz\'ow, ul. Pigonia 1, PL-35-310 Rzesz\'ow, Poland\vspace{1cm}}

\begin{abstract}
\vspace{0.5cm}
In this work, we present a thorough analysis of scalar $P$-wave $\chi_{Q0}$, $Q=c,b$ quarkonia electromagnetic form factors for the $\gamma^* \gamma^* \to \chi_{Q0}$ couplings, as well as their hadroproduction observables in $k_\perp$-factorisation using the light-front potential approach for the quarkonium wave function. The electromagnetic form factors are presented as functions of photon virtualities. We discuss the role of the Melosh spin-rotation and relativistic corrections estimated by comparing our results with those in the standard nonrelativistic QCD (NRQCD) approach. Theoretical uncertainties of our approach are found by performing the analysis with two different unintegrated gluon densities and with five distinct models of the $Q\bar Q$ interaction potentials consistent with the meson spectra. The results for the rapidity and transverse momentum distributions of $\chi_{Q0}$ produced in high-energy $pp$ collisions at $\sqrt{s} = 13 \, \rm{TeV}$ are shown.
\end{abstract}

\pacs{12.38.Bx, 13.85.Ni, 14.40.Pq}
\maketitle

\section{Introduction}

The production of quarkonia in hadronic collisions continues to attract a lot of 
interest \cite{Lansberg:2019adr,Chisholm:2014sca,Belyaev:2015lkw,Baranov:2019lhm}. Different approaches 
were used in the past, for the case of $\chi_c, \chi_b$ production, see e.g. \cite{Hagler:2000dd,Kniehl:2006sk,Baranov:2015yea,CS2018,Likhoded:2012hw,Likhoded:2014kfa,Diakonov:2012vb,Boer:2012bt}. 
Here, we continue our investigation \cite{BPSS2019,BGPSS2019} of quarkonia production observables in the framework  of the $k_\perp$-factorisation approach \cite{Gribov:1984tu,Catani:1990xk,Collins:1991ty} in the 
color-singlet formulation employing the light-front quarkonia wave functions. The previous works in Refs.~\cite{Hagler:2000dd,Kniehl:2006sk,Baranov:2015yea,CS2018} 
have observed that in the $k_\perp$-factorisation approach the color-singlet mechanism
dominates in the quarkonia production. At the next natural step we pursue in this direction,
it is instructive to explore the role of relativistic corrections and the shape of the
quarkonia wave function in the associated observables.

In our recent study \cite{BPSS2019,BGPSS2019} we discussed inclusive production of pseudoscalar $\eta_c(1S,2S)$ charmonia in proton-proton collisions. We explicitly calculated the
corresponding electromagnetic $\gamma^* + \gamma^* \to \eta_c(1S,2S)$ form factors as 
functions of photon virtualities. Similar form factors were calculated also for 
the $g^* + g^* \to \eta_c(1S,2S)$ matrix element that was used in the $k_T$-factorization
formula for differential hadroproduction cross section. We have shown there that inclusion 
of both form factors in such an approach is essential. In our calculations, we adopted the color-singlet model, which treats the quarkonium as a two-body bound state of heavy quark and antiquark. 

Such a formalism was used previously for the production of $\chi_{cJ}$ ($J=0,1,2$) quarkonia (see e.g.~Ref.~\cite{CS2018} and references therein), and a relatively good agreement with data was obtained in the $k_T$-factorization approach with an unintegrated gluon distribution (UGD), which effectively includes higher-order corrections. These previous calculations were done within the nonrelativistic QCD (NRQCD) approach.

In this work, we discuss production of the lowest $P$-wave scalar state for both $c\bar c$ and $b\bar b$ systems, $\chi_{Q0}$, $Q=c,b$, in the same approach which we dubbed at the light-front $k_\perp$-factorisation. As will be discussed in detail below, the situation with scalar $\chi_{c0}$ production is a bit more complicated than that for the pseudoscalar quarkonia. The leading-order partonic subprocess at high energies is the off-shell gluon-gluon fusion, $g^*+g^* \to \chi_{Q0}$. The main ingredient of such a subprocess is the off-shell matrix element written in terms of two independent form factors. Here we go beyond the NRQCD approach and present a similar analysis of $\chi_{Q0}$ form factors and hadroproduction as was done earlier for pseudoscalar ($S$-wave) charmonia. We will present the corresponding form factors, compare the new results to those obtained in NRQCD approaches and make predictions for the $p p \to \chi_{Q0}$ reactions at the LHC energies. 

The paper is organised as follows. In Sec.~\ref{sec:kt-fact}, we recapitulate the main details of the $k_\perp$-factorisation formalism. In Sec.~\ref{sec:off-shell-ME} we derive the off-shell matrix element expressed in terms of two independent form factors written in terms of light-front quarkonia wave functions. 
These form factors can be related to the standard 
electromagnetic transition form factors, which encode the helicity dependence of the virtual-photon fusion amplitudes. We derive the relevant 
expressions in Sec.~\ref{sec:gamma-gamma_FFs}. 
We then turn to numerical results, presenting the light front wave functions and radiative decay widths in Sec.~\ref{sec:LFWF_plots}. In Sec.~\ref{sec:EM_FF_plots} we present our results for the electromagnetic form factors. 
Our results for the transverse momentum and rapidity distributions of $\chi_{c0}$ and $\chi_{b0}$ mesons are shown
in Sec.~\ref{sec:hadroproduction}. In Sec.~\ref{sec:Conclusions} we conclude.

\section{$k_T$-factorization for inclusive hadroproduction}
\label{sec:kt-fact}

Let us briefly recapitulate the $k_T$-factorization formalism we use
in this work.
The leading process is the gluon-gluon fusion into the color-singlet heavy quark pair. 
Gluons carry transverse momenta, and their four momenta are written as
($\sqrt{s}$ is the $pp$ center-of-mass energy, and we use the lightcone components of four-vectors $a = (a_+,a_-,\ba)\, , \, a^2 = 2a_+a_- - \ba^2$):
\begin{eqnarray}
q_1 = (q_{1+},0,\bq_1) \, , \, q_2 = (0,q_{2-},
 \bq_2) \, ,
\end{eqnarray}
with 
\begin{eqnarray}
q_{1+} = x_1 \sqrt{s \over 2} \, , \, q_{2-} = x_2  \sqrt{s \over 2} \, .
\end{eqnarray}
An important distinction from the collinear approach is that 
the gluon momenta are (spacelike) off-shell, $q_i^2 = - \bq_i^2, i = 1,2$.
Our starting point is the inclusive cross section for the $2 \to 1$ gluon-gluon fusion mode obtained from
\begin{eqnarray}
d\sigma = \int {dx_1 \over x_1} \int {d^2 \bq_1 \over \pi \bq_1^2} 
{\cal{F}}(x_1,\bq_1^2,\mu_F^2)\int {dx_2 \over x_2} 
\int {d^2 \bq_2 \over \pi \bq_2^2}  {\cal{F}}(x_2,\bq_2^2,\mu_F^2) {1 \over 2 x_1 x_2 s} \overline{|{\cal{M}}|^2} \, d\Phi(2 \to 1). \nonumber \\
\end{eqnarray}
Here the unintegrated gluon distributions are normalized such that the collinear glue is given by
\begin{eqnarray}
xg(x,\mu_F^2) = \int^{\mu_F^2} {d \bq^2 \over \bq^2} {\cal{F}}(x,\bq^2,\mu_F^2) \, ,
\end{eqnarray}
where from now on we will no longer show the dependence on the factorization scale $\mu_F^2$ explicitly.
Let us denote the four-momentum of $\chi$ meson by $p$
and parametrize it in light-cone coordinates as
\begin{eqnarray}
p = (p_+,p_-,\bp) = \Big( {m_T \over \sqrt{2}} e^y, {m_T \over \sqrt{2}} e^{-y}, \bp \Big) \, , \qquad m_T = \sqrt{\bp^2 + M_\chi^2} .
\end{eqnarray}
Here $M_\chi$ is the mass of the $\chi_{Q0}$-meson, 
$\bp$ is its transverse momentum, and $y$ its rapidity in 
the $pp$ c.m. frame.
The phase-space element is
\begin{eqnarray}
d\Phi (2 \to 1) &=& (2 \pi)^4 \delta^{(4)}(q_1 + q_2 - p) \, {d^4 p \over (2 \pi)^3} \delta(p^2 - M_{\chi}^2) \nonumber \\
&=& 
{2 \pi \over s} \delta(x_1 - {m_T \over \sqrt{s}} e^y)\delta(x_2 - {m_T \over \sqrt{s}} e^{-y}) \delta^{(2)}(\bq_1 + \bq_2 - \bp) \,  dy \, d^2\bp \, .
\end{eqnarray}
We therefore obtain for the inclusive cross section
\begin{eqnarray}
{d \sigma \over dy d^2\bp} = \int {d^2 \bq_1 \over \pi \bq_1^2} 
{\cal{F}}(x_1,\bq_1^2) \int {d^2 \bq_2 \over \pi \bq_2^2}  {\cal{F}}(x_2,\bq_2^2) \, \delta^{(2)} (\bq_1 + \bq_2 - \bp ) \, {\pi \over (x_1 x_2 s)^2} \overline{|{\cal{M}}|^2} ,
\end{eqnarray}
where the momentum fractions of gluons are fixed as $x_{1,2} = m_T \exp(\pm y) / \sqrt{s}$.
The off-shell matrix element is written in terms of the Feynman amplitude
as (we restore the color-indices):
\begin{eqnarray}
{\cal{M}}^{ab} = {q_{1 \perp}^\mu q_{2\perp}^\nu \over |\bq_1| |\bq_2|}{\cal{M}}^{ab}_{\mu \nu}  = {q_{1+} q_{2-} \over |\bq_1| |\bq_2|} n^+_\mu n^-_\nu {\cal{M}}^{ab}_{\mu \nu} = {x_1 x_2 s \over 2 |\bq_1| |\bq_2| } n^+_\mu n^-_\nu {\cal{M}}^{ab}_{\mu \nu}  \, .
\end{eqnarray}

\section{$k_T$-factorization element for final state 
$Q \bar Q$}
\label{sec:off-shell-ME}
Let us introduce the reduced amplitude
\begin{eqnarray}
 {\cal T}_{\mu \nu} \equiv {1 \over 4 \pi \alpha_{\rm em} e_Q^2} \, \int {dz d^2\bk \over z(1-z) 16 \pi^3} 
\sum_{\lambda, \bar \lambda} 
\Psi^*_{\lambda \bar \lambda}(z,\bk) 
{\cal M}_{\mu \nu}^{\lambda \bar \lambda} (\gamma^* \gamma^* \to Q_\lambda(zP_+,\bp_Q) \bar Q_{\bar \lambda} ((1-z)P_+,\bp_{\bar Q}) \, . \nonumber \\
\end{eqnarray}
Here the transverse momenta of quark and antiquark are
\begin{eqnarray}
\bp_Q = \bk + z \bP \, , \qquad \bp_{\bar Q} = - \bk + (1-z) \bP \, .
\end{eqnarray}
The photon-photon fusion and gluon-gluon fusion amplitudes are proportional to the reduced amplitude
as follows:
\begin{eqnarray}
{\cal M}_{\mu \nu} (\gamma^* \gamma^* \to \chi_{Q0}(P_+,\bP)) &=& 4 \pi \alpha_{\rm em} e_Q^2 {\Tr[\openone] \over \sqrt{N_c}} \,  {\cal T}_{\mu \nu}
= 4 \pi \alpha_{\rm em} e_Q^2 \sqrt{N_c} \,  {\cal T}_{\mu \nu} \, ,
\nonumber \\
{\cal M}^{ab}_{\mu \nu} (g^* g^* \to \chi_{Q0}(P_+,\bP)) &=& 4 \pi \alpha_s {\Tr[t^a t^b] \over \sqrt{N_c}} \,  {\cal T}_{\mu \nu} = {2 \pi \alpha_s \over \sqrt{N_c}} \, \delta^{ab} {\cal T}_{\mu \nu} \, .
\end{eqnarray}
For the purpose of $k_T$-factorization it is most convenient to use the contraction of the gluon-fusion amplitude with light-cone vectors $n_\mu^{\pm}$. 
Using the standard covariant Feynman rules, we can obtain the reduced amplitude ${\cal T} = n^{+\mu} n^{-\nu} {\cal T}_{\mu \nu}$ as
\begin{eqnarray}
{\cal T} &=& (-2) \int {dz d^2\bk \over \sqrt{z(1-z)} 16 \pi^3}
\nonumber \\
&\times& \Big\{ -m_Q \Big[ {1 \over \bl_A^2 + \varepsilon^2} - {1 \over \bl_B^2 + \varepsilon^2}\Big]
\Big( \sqrt{2}(\be(-)\bq_1) \Psi^*_{++}(z,\bk) + \sqrt{2} (\be(+)\bq_1) \Psi^*_{--}(z,\bk) \Big) \nonumber \\
&+& \Big( 2 z(1-z) \bq_1^2 + (1-2z) (\bk \bq_1) \Big) \Big[ {1 \over \bl_A^2 + \varepsilon^2} - {1 \over \bl_B^2 + \varepsilon^2}\Big] \Big( \Psi^*_{+-}(z,\bk) 
+ \Psi^*_{-+}(z,\bk) \Big) \nonumber \\
&-& (1-2z) (\bq_1 \bq_2) \Big[ {1-z \over \bl_A^2 + \varepsilon^2} + {z \over \bl_B^2 + \varepsilon^2} \Big] \Big( \Psi^*_{+-}(z,\bk) 
+ \Psi^*_{-+}(z,\bk) \Big) \Big\} \; .
\end{eqnarray}
Here we have introduced the shorthand notation
\begin{eqnarray}
\varepsilon^2 &=& z(1-z)\bq_1^2 +m_Q^2\,,
\qquad \bl_A = \bk - (1-z) \bq_2\,, \qquad 
\bl_B = \bk + z \bq_2 \, .
\end{eqnarray}
We have also made use of the fact, that for the scalar meson
\begin{eqnarray}
 \Psi^*_{+-}(z,\bk) - \Psi^*_{-+}(z,\bk) &=& 0.
\end{eqnarray}
We have collected the necessary information on the light-front 
wave function $\Psi_{\lambda \bar \lambda}(z,\bk)$ in Appendix \ref{sec:LCWF}. 
Inserting the identities which follow from Eq.(\ref{eq:Psi_matrix}): 
\begin{eqnarray}
\sqrt{2} (\be(-)\bq_1) \Psi^*_{++}(z,\bk)
+\sqrt{2} (\be(+)\bq_1) \Psi^*_{--}(z,\bk)
&=& {2 (\bq_1 \bk) \over \sqrt{z(1-z)}}  \psi(z,\bk) \, , \nonumber \\
\Psi^*_{+-}(z,\bk) + \Psi^*_{-+}(z,\bk) &=& {2m_Q (1-2z) \over \sqrt{z(1-z)}}  \psi(z,\bk) \, ,
\end{eqnarray}
we can simplify the amplitude further to read
\begin{eqnarray}
{\cal T} 
&=& (-4m_Q) \int {dz d^2\bk \over z(1-z) 16 \pi^3} \psi(z,\bk)
\Big\{
2 z(1-z) (1-2z) \bq_1^2 \Big[ {1 \over \bl_A^2 + \varepsilon^2} - {1 \over \bl_B^2 + \varepsilon^2}\Big] \nonumber \\
&-& 4z(1-z) (\bk \bq_1) \Big[ {1 \over \bl_A^2 + \varepsilon^2} - {1 \over \bl_B^2 + \varepsilon^2}\Big] 
- (1-2z)^2 (\bq_1 \bq_2) \Big[ {1-z \over \bl_A^2 + \varepsilon^2} + {z \over \bl_B^2 + \varepsilon^2} \Big] 
\Big\} \; . 
\end{eqnarray}
There now emerge two independent form factors $G_{1,2}(\bq_1^2,\bq_2^2)$:
\begin{eqnarray}
 {\cal T} = |\bq_1||\bq_2| G_1(\bq_1^2,\bq_2^2) + (\bq_1 \cdot \bq_2) G_2(\bq_1^2,\bq_2^2) \, .
 \label{eq:FF_G1G2}
\end{eqnarray}
The form factors $G_{1,2}(\bq_1^2, \bq_2^2)$ have the integral representations
\begin{eqnarray}
 G_1(\bq_1^2, \bq_2^2) &=& |\bq_1||\bq_2| \,  {4m_Q \over \bq_2^2}  \int {dz d^2\bk \over z(1-z) 16 \pi^3} \psi(z,\bk) \,  2z(1-z) (2z-1) \Big[ {1 \over \bl_A^2 + \varepsilon^2} - {1 \over \bl_B^2 + \varepsilon^2}\Big] \nonumber \\
 G_2(\bq_1^2,\bq_2^2) &=&
 4m_Q \int {dz d^2\bk \over z(1-z) 16 \pi^3} \psi(z,\bk) \Big[ {1 -z \over \bl_A^2 + \varepsilon^2} + {z \over \bl_B^2 + \varepsilon^2}\Big] \nonumber \\
 &+& {4 m_Q \over \bq_2^2} \int {dz d^2\bk \over z(1-z) 16 \pi^3} \psi(z,\bk) 4z(1-z) 
 \Big[ {\bq_2\cdot \bl_A \over \bl_A^2 + \varepsilon^2} - {\bq_2 \cdot \bl_B \over \bl_B^2 + \varepsilon^2}\Big] \, .
 \label{eq:G1_G2}
\end{eqnarray}
We explain in Appendix \ref{sec:LCWF} how the light-front radial WF $\psi(z,\bk)$ can be obtained from the potential model WF using the Melosh-transform technique for the spin-orbit part \cite{Melosh:1974cu,Jaus:1989au} augmented by the Terent'ev prescription \cite{Terentev:1976jk}.

Finally, the factorization formula for the hadroproduction process
$pp \to \chi_{Q0} X$ reads as follows
\begin{eqnarray}
{d \sigma \over dy d^2 \bp} &=& \int {d^2 \bq_1 \over \pi \bq_1^4} { d^2 \bq_2 \over \pi \bq_2^4} 
\delta^{(2)}(\bp - \bq_1 -\bq_2)
{\cal F}(x_1,\bq_1) {\cal F}(x_2,\bq_2)
 {\pi \over 4 (N_c^2-1)^2} \, \sum_{a,b} 
|{\cal M}^{ab}_{\mu \nu}n^+_\mu n^-_\nu|^2 \, \nonumber \\
&=&  \int {d^2 \bq_1 \over \pi \bq_1^4} { d^2 \bq_2 \over \pi \bq_2^4} 
\delta^{(2)}(\bp - \bq_1 -\bq_2)
{\cal F}(x_1,\bq_1) {\cal F}(x_2,\bq_2)
{\pi^3 \alpha_S^2 |{\cal T}|^2\over N_c(N_c^2-1)} \; \nonumber \\
&=&  \int {d^2 \bq_1 \over \pi \bq_1^2} { d^2 \bq_2 \over \pi \bq_2^2} 
\delta^{(2)}(\bp - \bq_1 -\bq_2)
{\cal F}(x_1,\bq_1) {\cal F}(x_2,\bq_2)
{\pi^3 \alpha_S^2 \over N_c(N_c^2-1)} 
\nonumber \\
&\times& 
\Big( G_1(\bq_1^2,\bq_2^2) + \cos(\phi_1 - \phi_2) G_2(\bq_1^2,\bq_2^2) \Big)^2\; , 
\end{eqnarray}
with $\phi_{1,2}$ being the azimuthal angles of gluon
transverse momenta $\bq_{1,2}$.
In our numerical calculations presented below, we set the factorization
scale to $\mu_F^2 = m_T^2$, and the QCD coupling constant squared in the differential cross section above is taken in the following form:
\begin{equation}
\alpha_s^2 \to 
\alpha_s(\max{\{m_T^2,\bq_1^2\}})
\alpha_s(\max{\{m_T^2,\bq_2^2\}})  \; .
\label{alpha_s}
\end{equation}
%

\subsection{NRQCD limit}

Let us now take the nonrelativistic (NR) limit.
In this case, one should Taylor-expand around $z = 1/2$ and $\bk =0$.
Let us introduce
\begin{eqnarray}
z = \half + \xi \, , \qquad
1-z = \half - \xi \, .
\end{eqnarray}
We can see, that in zeroth order the form factors $G_1, G_2$ vanish.
To the first order in $\xi$ and $\bk$, we obtain
\begin{eqnarray}
{1\over \bl_A^2 + \varepsilon^2} - 
{1\over \bl_B^2 + \varepsilon^2} = {2 \over \mu^4} \Big( \xi \bq_2^2 + (\bk \bq_2) \Big) \, ,
\end{eqnarray}
with
\begin{eqnarray}
\mu^2 = {1 \over 4} \Big( \bq_1^2 + \bq_2^2 + 4 m_c^2 \Big) \, .
\end{eqnarray}
In the NR limit, we also replace $2m_Q \to M_\chi$,
and obtain for the two form factors
\begin{eqnarray}
 G_1(\bq_1^2, \bq_2^2) &=& {|\bq_1||\bq_2| \over \mu^4} {4 \over M_\chi} \,  \int {dz d^2\bk \over z(1-z) 16 \pi^3} \psi(z,\bk) \, \xi^2 M_\chi^2
 \, , \nonumber \\
 G_2(\bq_1^2, \bq_2^2) &=& {2 \over \mu^4} {1 \over M_\chi} \int {dz d^2\bk \over z(1-z) 16 \pi^3} \psi(z,\bk) \Big( \bk^2 M_\chi^2 + 4 \xi^2 M_\chi^2 \mu^2 \Big) \, .
\end{eqnarray}
Now, again in the NR limit, we can replace $\xi M_\chi = k_z$, and substitute
\begin{eqnarray}
 {dz d^2\bk \over z(1-z) 16 \pi^3} \psi(z,\bk)  \to {1 \over 4 \pi^2 \sqrt{M_\chi}} {1 \over 2 \sqrt{2}} d^3k {u(k) \over k^2} \, .
 \label{eq:psi_to_uk}
\end{eqnarray}
Using the relation 
\begin{eqnarray}
 \int_0^\infty dk \, k^2 u(k) = 3  \sqrt{\pi \over 2} R'(0)\, ,
\label{eq:uk_to_R0}
\end{eqnarray}
we see that both form factors are proportional to the first derivative of the radial wave function at the origin $R'(0)$:
\begin{eqnarray}
 G_1(\bq_1^2, \bq_2^2) &=& {2 |\bq_1||\bq_2| \over \
 [ M_\chi^2 + \bq_1^2 + \bq_2^2]^2 } \, \lambda \, , \qquad
 G_2(\bq_1^2,\bq_2^2) = {3 M_\chi^2 + \bq_1^2 + \bq_2^2 \over [ M_\chi^2 + \bq_1^2 + \bq_2^2]^2} \, \lambda \, .
\end{eqnarray}
Here we introduced the short-hand notation
\begin{eqnarray}
 \lambda = {8 \over \sqrt{\pi}} {R'(0) \over M^{3/2}} \, .
\end{eqnarray}
The reduced amplitude ${\cal T}$ now becomes:
\begin{eqnarray}
{\cal T} 
&=&  {\lambda \over [M_{\chi}^2 + \bq_1^2 + \bq_2^2]^2 } \,
\Big( 2 \bq_1^2 \bq_2^2 + (\bq_1 \bq_2) (3  M_{\chi}^2 + \bq_1^2 + \bq_2^2) \Big)\, .
\label{eq:tau_NRQCD}
\end{eqnarray}
This result is in full agreement with Refs. 
\cite{Kniehl:2006sk,Pasechnik:2007hm}.

\section{The $\gamma^* \gamma^* \to \chi_{Q0}$ amplitude and electromagnetic transition form factors}
\label{sec:gamma-gamma_FFs}

The photon-photon fusion amplitude can be expressed in terms of two form-factors $F_{TT}, F_{LL}$, referring
to photons having transverse or longitudinal polarizations in the $\gamma^* \gamma^*$ c.m.s. with the quantization axis taken along the $\gamma^* \gamma^*$ collision axis.
Using the standard techniques reviewed in \cite{Budnev:1974de, Poppe:1986dq}, it can be written covariantly as
\begin{eqnarray}
 { 1 \over 4 \pi \alpha_{\rm em}} {\cal M}_{\mu \nu} (\gamma^*(\bq_1) \gamma^*(\bq_2) \to \chi_{Q0}(\bP)) = - \delta^\perp_{\mu \nu} (q_1,q_2) \, F_{TT}(q_1^2,q_2^2) 
 + e_\mu^L(q_1) e_\nu^L(q_2) \, F_{LL}(q_1^2, q_2^2) \, . \nonumber \\
 \label{eq:gammagamma_amplitude}
\end{eqnarray}
Here we have introduced the projector on transverse polarization states
\begin{eqnarray}
 - \delta^\perp_{\mu \nu} (q_1,q_2) = - g_{\mu \nu} + {1 \over X} \Big( (q_1 \cdot q_2)(q_{1\mu} q_{2\nu} + q_{1 \nu} q_{2\mu}) 
 - q_1^2 q_{2\mu} q_{2\nu}- q_2^2 q_{1\mu} q_{1\nu} \Big) \, ,
\end{eqnarray}
with $X = (q_1 \cdot q_2)^2 - q_1^2 q_2^2$.
The longitudinal polarization states of virtual photons read:
\begin{eqnarray}
 e_\mu^L(q_1) &=& \sqrt{ {-q_1^2 \over X}} \Big (q_{2\mu} - { q_1 \cdot q_2 \over q_1^2} q_{1\mu} \Big)\, , \qquad
 e_\nu^L(q_2) = \sqrt{ {-q_2^2 \over X}} \Big (q_{1\nu} - { q_1 \cdot q_2 \over q_2^2} q_{2\nu} \Big)\, .
\end{eqnarray}
Notice that here we always consider polarization in the 
$\gamma \gamma$-c.m. frame.
From here, we can obtain the decay width for $S \to \gamma \gamma$ as
\begin{eqnarray}
 \Gamma(\chi_{Q0} \to \gamma \gamma) = { \pi \alpha_{\rm em}^2 \over M_\chi} |F_{TT}(0,0)|^2 \, ,
 \label{eq:Gamma}
\end{eqnarray}
where $M_\chi$ is the mass of the scalar meson.

We now wish to relate the helicity form factors $F_{TT}, F_{LL}$ to the form factors of the (reggeized) gluon fusion amplitude, $G_1,G_2$.
We start by evaluating the contractions of off-shell ``polarizations'' $n^{\pm}_\mu$ with the projectors on transverse and longitudinal helicities:
\begin{eqnarray}
-n^{+\mu} n^{-\nu} \delta^\perp_{\mu\nu} &=& -1 + {(q_1\cdot q_2) \over X} q_1^+ q_2^- = {1 \over X} \Big( q_1^2 q_2^2 + (\bq_1 \cdot \bq_2) (q_1 \cdot q_2) \Big) \,,
\end{eqnarray}
and
\begin{eqnarray}
e_\mu^L(q_1) e_\nu^L(q_2) n^{+\mu} n^{-\nu} &=& {\sqrt{q_1^2 q_2^2} \over X} q_1^+ q_2^-
=  {\sqrt{q_1^2 q_2^2} \over X}\Big( (q_1 \cdot q_2) + (\bq_1 \cdot \bq_2) \Big) \,.
\end{eqnarray}
Inserting these projections into the general form of the amplitude and collecting coefficients in front of $\bq_1 \cdot \bq_2$ and $q_1^2 q_2^2$, we obtain
\begin{eqnarray}
{n^{+\mu} n^{-\nu} \over 4 \pi \alpha_{\rm em}} {\cal M}_{\mu \nu} &=&
(\bq_1 \cdot \bq_2) \Big[{(q_1 \cdot q_2) \over X} F_{TT}(q_1^2,q_2^2) + {|\bq_1||\bq_2| \over X} F_{LL}(q_1^2, q_2^2)\Big] \nonumber \\
&+& |\bq_1||\bq_2| \Big[ {|\bq_1||\bq_2| \over X} F_{TT}(q_1^2,q_2^2) + {(q_1 \cdot q_2) \over X} F_{LL}(q_1^2, q_2^2)\Big] \, .
\end{eqnarray}
Here we used, that $\sqrt{q_1^2 q_2^2} = |\bq_1||\bq_2|$.
Let us also note, that
\begin{eqnarray}
(q_1 \cdot q_2) &=& {1 \over 2}\Big( M_\chi^2 + \bq_1^2 + \bq_2^2 \Big) 
\; {\rm and} \; 
X = {M_\chi^4 \over 4} \Big( 1 + {2 (\bq_1^2 + \bq_2^2) \over M_\chi^2} + {(\bq_1^2 - \bq_2^2)^2 \over M_\chi^4} \Big) \, , \nonumber
\end{eqnarray}
do not depend on the azimuthal angles of $\bq_1$ and $\bq_2$.
Thus we can now easily read off the relation between the two sets of form factors:
\begin{eqnarray}
\left( \begin{array}{c} G_1 \\  G_2 \end{array} \right) = { 1 \over e_Q^2 \sqrt{N_c} X} 
\begin{pmatrix} |\bq_1||\bq_2| & (q_1 \cdot q_2) \\ (q_1\cdot q_2) &
  |\bq_1||\bq_2| \end{pmatrix} \; \left( \begin{array}{c} 
    F_{TT} \\ F_{LL} \end{array} \right) \; .
\end{eqnarray}
This set of linear equations can be easily inverted, so that helicity form factors can be calculated using the form factors $G_{1,2}$:
\begin{eqnarray}
\left( \begin{array}{c} F_{TT} \\ F_{LL} \end{array} \right) &=& e_Q^2 \sqrt{N_c} 
\begin{pmatrix} - |\bq_1||\bq_2| & (q_1 \cdot q_2) \\  (q_1\cdot q_2) &
  - |\bq_1||\bq_2| \end{pmatrix} \; \left( \begin{array}{c} 
    G_1 \\  G_2 \end{array} \right) \; . \nonumber \\
 \label{eq:FTT_FLL}
\end{eqnarray}
At the on-shell point, $\bq_i^2=0$, we have 
\begin{eqnarray}
F_{TT}(0,0) = {M_\chi^2 \over 2} \, G_2(0,0) \, ,
\end{eqnarray}
while $G_1(0,0) = 0$.
In the NRQCD limit, we obtain
\begin{eqnarray}
 F_{TT}(0,0) = e_Q^2\sqrt{N_{c}}{12 \over \sqrt{\pi}}{R'(0)\over M_\chi^{3/2}} \, ,
\end{eqnarray}
which leads to the known relation of the radiative decay width to $R'(0)$:
\begin{eqnarray}
\Gamma(\chi_{Q0} \to \gamma\gamma) &=& \frac{9\cdot2^4\alpha_{\rm em}^2
   e_{Q}^{4} N_{c}}{M^4_{\chi}} |R'(0)|^2 \;.
   \label{eq:Gamma_R0_LO}
\end{eqnarray}
We can also decompose form factors $G_1,G_2$ into $TT$ and $LL$ components. 
Indeed, the reduced amplitude ${\cal T}$ has the decomposition analogous to Eq.(\ref{eq:gammagamma_amplitude}):
\begin{eqnarray}
 {\cal T}_{\mu \nu}
 = - \delta^\perp_{\mu \nu} (q_1,q_2) \, {F_{TT}(q_1^2,q_2^2) \over e_Q^2 \sqrt{N_c}}
 + e_\mu^L(q_1) e_\nu^L(q_2) \, {F_{LL}(q_1^2, q_2^2) \over e_Q^2 \sqrt{N_c}} \, , 
 \label{eq:gluongluon_amplitude}
\end{eqnarray}
and hence
\begin{eqnarray}
 {\cal T} = n^{+\mu}n^{-\nu} {\cal T}_{\mu \nu} = {\cal T}_{TT} + {\cal T}_{LL} \, .
\end{eqnarray}
Evidently, each of the amplitudes ${\cal T}_{TT}, {\cal T}_{LL}$ has a decomposition like in Eq.(\ref{eq:FF_G1G2}):
\begin{eqnarray}
 {\cal T}_{TT} &=& |\bq_1||\bq_2| G_{1,TT}(\bq_1^2,\bq_2^2) + (\bq_1 \cdot \bq_2) G_{2,TT}(\bq_1^2,\bq_2^2) \, ,\nonumber \\
 {\cal T}_{LL} &=& |\bq_1||\bq_2| G_{1,LL}(\bq_1^2,\bq_2^2) + (\bq_1 \cdot \bq_2) G_{2,LL}(\bq_1^2,\bq_2^2) \, ,
\end{eqnarray}
with
\begin{eqnarray}
 G_{1,TT} &=& {|\bq_1||\bq_2| \over X} \Big( 
 \half (M_\chi^2+ \bq_1^2 + \bq_2^2) G_2 - |\bq_1||\bq_2| G_1 \Big) \, , \nonumber \\
 G_{2,TT} &=& {M_\chi^2 + \bq_1^2 + \bq_2^2 \over 2 X} \Big( 
 \half (M_\chi^2+ \bq_1^2 + \bq_2^2) G_2 - |\bq_1||\bq_2| G_1 \Big) \, , \nonumber \\
 G_{1,LL} &=& {M_\chi^2 + \bq_1^2 + \bq_2^2 \over 2 X} \Big( 
 \half (M_\chi^2+ \bq_1^2 + \bq_2^2) G_1 - |\bq_1||\bq_2| G_2 \Big) \, , \nonumber \\
 G_{2,LL} &=& {|\bq_1||\bq_2| \over X} \Big( 
 \half (M_\chi^2+ \bq_1^2 + \bq_2^2) G_1 - |\bq_1||\bq_2| G_2 \Big) \, .
 \label{eq:G_TT_G_LL}
\end{eqnarray}
This decomposition puts into evidence that the fusion of two off-shell (reggeized) gluons contains transverse as well as longitudinally polarized gluons in the 
frame of the produced system. The longitudinal polarizations are absent in approaches based on on-shell gluons.

\section{Light-front wave functions and radiative decay widths}
\label{sec:LFWF_plots}

\begin{figure}
    \centering
    \includegraphics[width=0.45\linewidth]{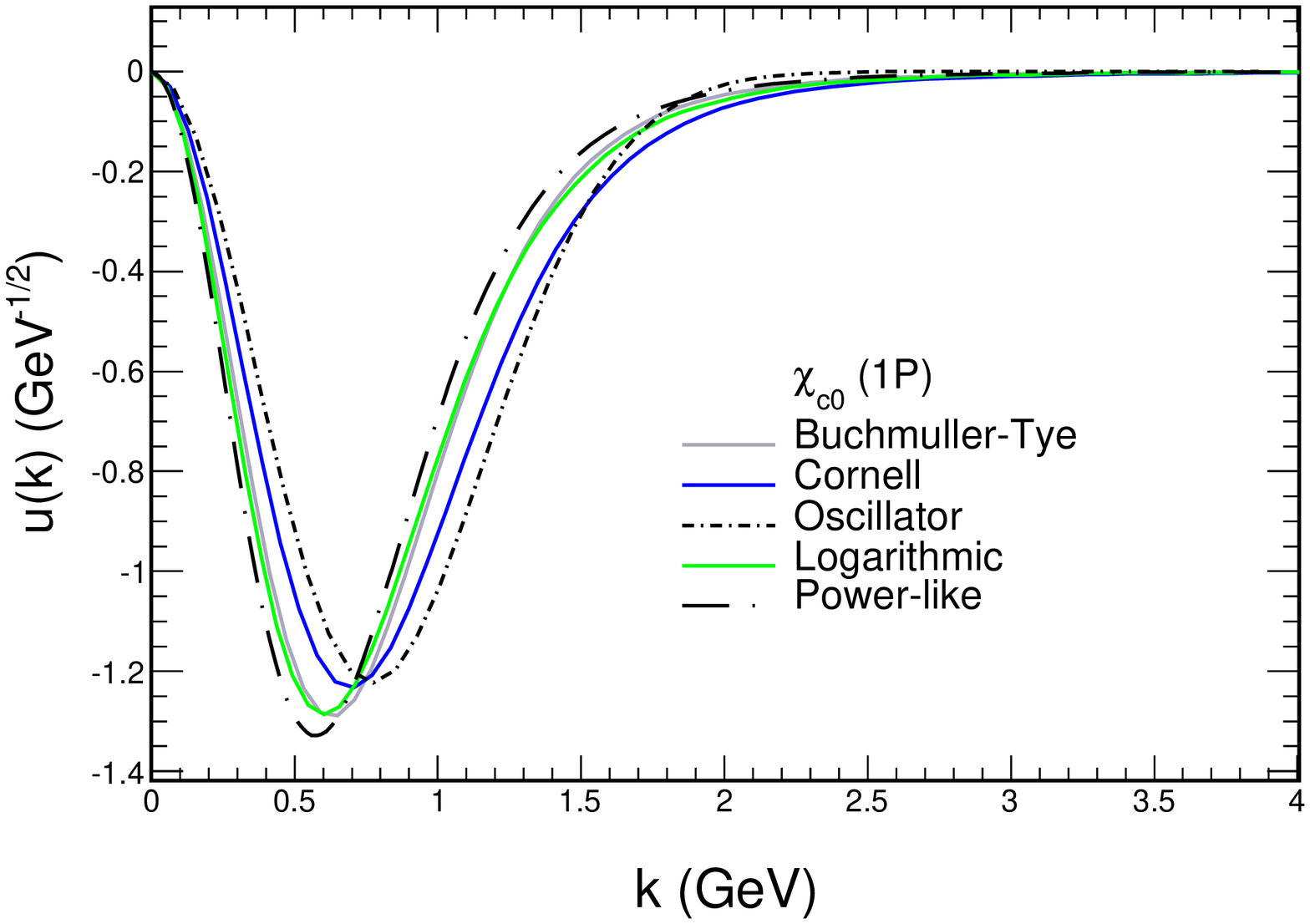}
     \includegraphics[width=0.45\linewidth]{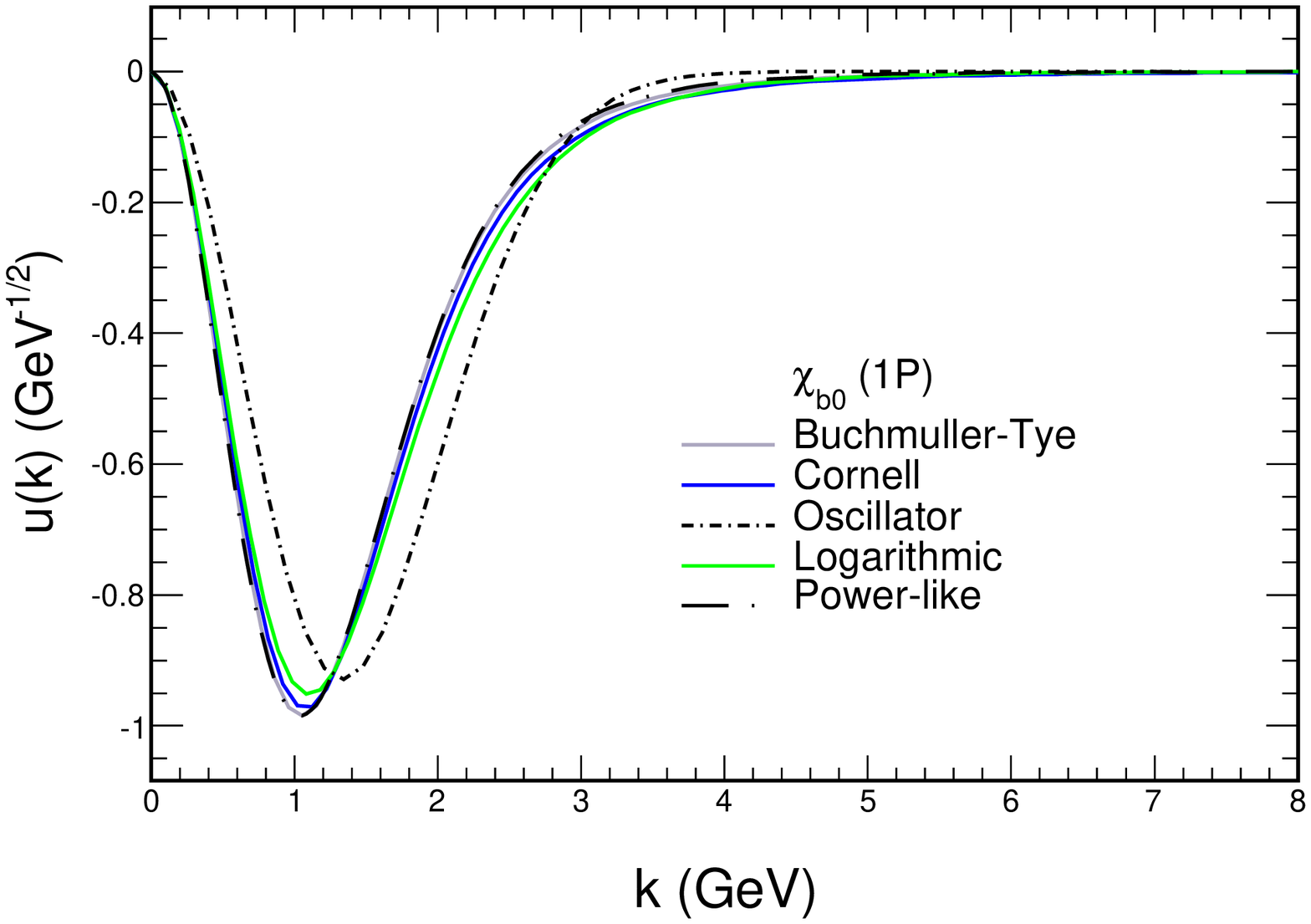}
    \caption{Quarkonium wave functions $u(k)$ for the $\chi_{c0}$ and the $\chi_{b0}$ for five potentials.}
    \label{fig:up}
\end{figure}

In Fig.\ref{fig:up} we show the nonrelativistic wave function $u(k)$ obtained
with different $c \bar c$ potentials from the literature. We refer the reader
to \cite{BGPSS2019,Hufner:2000jb} where more details can be found.
The P-wave functions have extra zero at $k=0$ compared to the S-wave function
discussed in \cite{BGPSS2019}. 
This will have important consequences for corresponding transition form factors.

\begin{figure}
    \centering
    \includegraphics[width=0.45\linewidth]{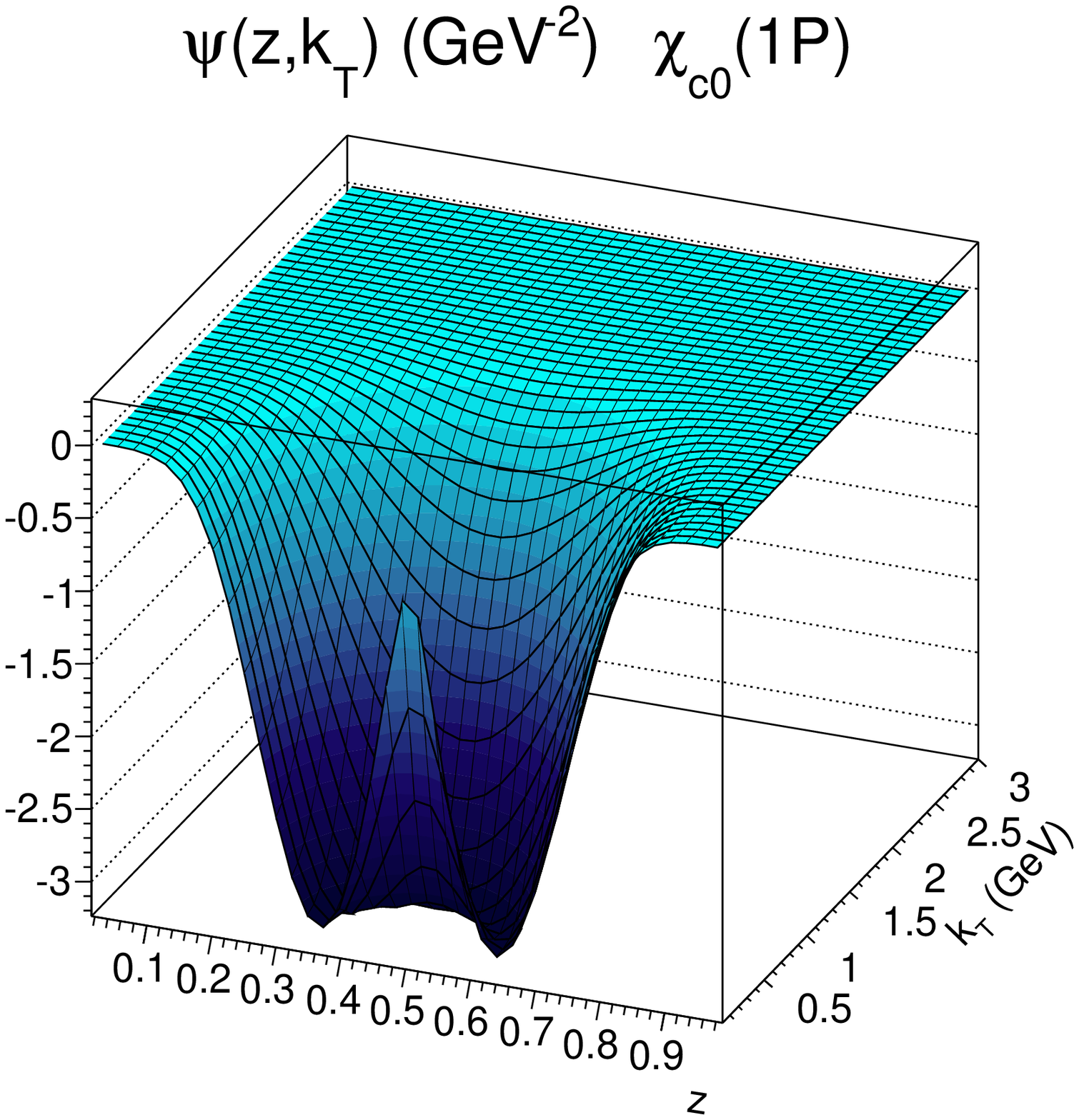}
    \includegraphics[width=0.45\linewidth]{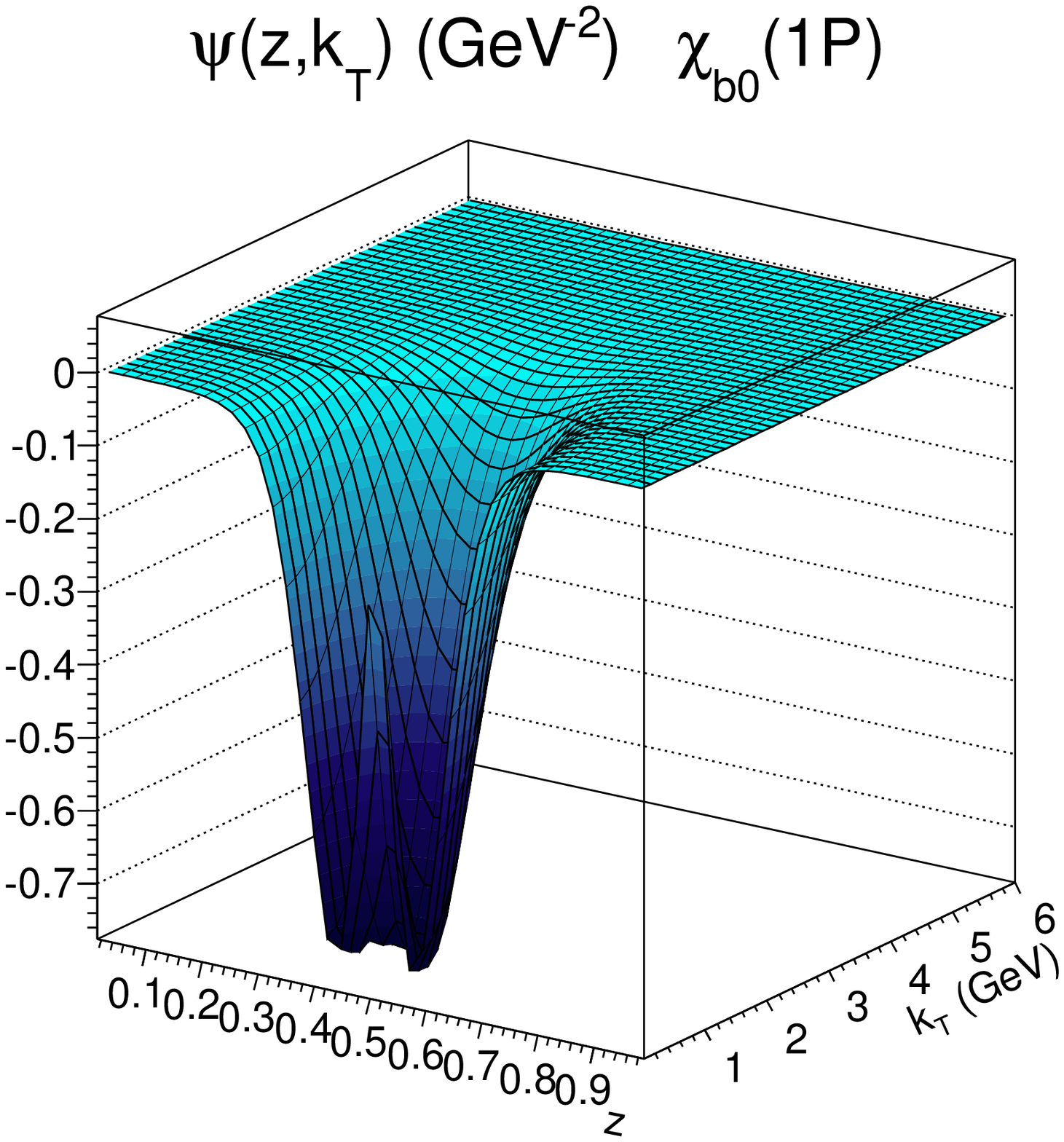}
    \caption{Light cone wave function $\psi(z,k_{T})$ for 
            the Buchm\"uller-Tye potential for $\chi_{c0}$ and $\chi_{b0}$.}
    \label{fig:LFWF_but}
\end{figure}

The corresponding light-cone wave function is shown for example in
Fig.\ref{fig:LFWF_but} for the Buchm\"uller-Tye potential.
Similar light-front wave functions were obtained from four
other potentials from the literature. The wave functions are used then
to calculate $\gamma^* \gamma^* \to \chi_{c0}$ and $\gamma^* \gamma^* \to \chi_{b0}$  form factors as discussed in the previous section.
For both mesons the light-cone wave function has a similar shape but in the case of the $\chi_{b0}$ the range of $k_{T}$ is broader.

Some basic information on $\chi_{c0}$ and $\chi_{b0}$ mesons is collected in Tab.~\ref{tab:table1}.
\begin{table}[h!]
    \centering
        \caption{Values for meson mass, $\Gamma_{tot}$, and the branching ratio ${\rm Br}(\chi_{Q0} \to \gamma \gamma) = \Gamma(\chi_{Q0} \to \gamma \gamma)/\Gamma_{tot}$
         according to the Particle Data
          Group \cite{PDG}.}
    \begin{tabular}{c|c|c|c}
    \hline
    \hline
                      & $M_{\chi}(1P) [MeV]$ &$\Gamma_{tot} [MeV]$ &${\rm Br}(\chi_{Q0} \to \gamma \gamma)$\\
                      \hline
     $\chi_{c0}(1P)$ &  $3414.71\pm 0.30$&$10.8\pm0.6$ & $(2.04\pm0.09)\cdot10^{-4}$ \\
       $\chi_{b0}(1P)$ & 9859.44$\pm$0.42$\pm$0.31 & - &  - \\
      \hline
      \hline
    \end{tabular}
    \label{tab:table1}
\end{table}
The value of $|F_{TT}(0,0)|$ and decay widths at the leading and next-to-leading order obtained from different potential models are collected in Tab.~\ref{tab:table2} for $\chi_{c0}$ meson. 
We used the NLO expression for the radiative decay width (see Ref.~\cite{Ebert:2003mu}), inserting the LO relation between $\gamma\gamma$ decay width and $F(0,0)$ from Eq.~(\ref{eq:Gamma}):
\begin{eqnarray}
\Gamma(\chi_{Q0} \to \gamma \gamma)_{NLO} &=& \Gamma(\chi_{Q0} \to \gamma \gamma)_{LO}
\Big{[}1 + \frac{\alpha_{s}}{\pi}\Big{(}\frac{\pi^2}{3}-
\frac{28}{9}\Big{)}\Big{]}  \; ,
\label{eq:Gamm_NLO}
\end{eqnarray}
here we employed $\alpha_s = 0.26$ for $\chi_{c0}$ and $\alpha_{s} = 0.18$ for $\chi_{b0}$ (see Ref.~\cite{Ebert:2003mu}).

Using Eq.~(\ref{eq:Gamm_NLO}) and the experimental value for the  radiative decay width we extracted $F_{TT}(0,0)$ for both on-shell photons at leading order.

\begin{table}[h!]
    \centering
        \caption{$F_{TT}(0,0)$ and $\Gamma(\chi\rightarrow\gamma\gamma)$
          for $\chi_{c0}$ for five different potentials.
        The value marked by * is obtained from Eq.~(\ref{eq:Gamma}), (\ref{eq:Gamm_NLO}). }
    \begin{tabular}{l|c|c|c}
    \hline
    \hline
    Potential type & $|F(0,0)|[GeV]$ & $\Gamma(\chi\rightarrow\gamma\gamma)_{LO} [keV]$ &  $\Gamma(\chi\rightarrow\gamma\gamma)_{NLO} [keV]$\\
    \hline
    Harmonic oscillator & 0.18 & 1.56 & 1.58\\
    Logarithmic         & 0.14 & 0.91 & 0.93\\
    Powerlike           & 0.16 & 1.32 & 1.34\\
    Cornell             & 0.10 & 0.44 & 0.46\\
    Buchm\"ulller-Tye   & 0.14 & 0.96 & 0.98\\
    \hline
    extracted from experiment \cite{PDG} & 0.21*& &$2.20\pm 0.16$\\
    \hline
    \hline
    \end{tabular}
    \label{tab:table2}
\end{table}

Similarly $|F_{TT}(0,0)|$, but with the PDG value of $m_c$ \cite{PDG} in the form factor is presented in Tab.~\ref{tab:table3}.
It is worth to notice, that the form factor is very sensitive to the quark mass.
This feature of the form factor is manifested in Tabs.~\ref{tab:table2}, \ref{tab:table3}
and Fig.~\ref{fig:FF_F0}.

\begin{table}[]
    \centering
    \caption{$F_{TT}(0,0)$ with $m_c = 1.27 \, \rm{GeV}$ and corresponding
      radiative widths.}
    \begin{tabular}{l|c|c|c}
    \hline
    \hline
    Potential type & $|F_{TT}(0,0)|[GeV]$ & $\Gamma(\chi_{c0}\rightarrow\gamma\gamma)_{LO} [keV]$ &  $\Gamma(\chi_{c0}\rightarrow\gamma\gamma)_{NLO} [keV]$\\
    \hline
    Harmonic oscillator  & 0.21 & 2.06 & 2.09 \\
    Logarithmic          & 0.18 & 1.54 & 1.56\\
    Power-law            & 0.18 & 1.54 & 1.56\\
    Cornell              & 0.17 & 1.41 & 1.43\\
    Buchm\"ulller-Tye    & 0.18 & 1.54 & 1.56\\
    \hline
    extracted from experiment \cite{PDG}& 0.21*& &$2.20\pm 0.16$\\
    \hline
    \hline
    \end{tabular}
    
    \label{tab:table3}
\end{table}
We apply the same formalism also to the $\chi_{b0}$ meson. The corresponding results are 
summed up in Tabs.~\ref{tab:FF_00_chib} and \ref{tab:FF_00_chib_mb}, respectively for the potential model $m_{b}$ value and for the PDG $m_{b}$ value.
\begin{table}[]
    \centering
        \caption{
        $|F_{TT}(0,0)|$ for $\chi_{b0}$ with quark/antiquark mass corresponding to each potential. Here $\alpha_{s}= 0.18$ according to Ref.~\cite{Ebert:2003mu}.}
    \begin{tabular}{l|c|c|c|c}
    \hline
    \hline
    Potential type  & $m_{b} [GeV]$ & $|F_{TT}(0,0)|[GeV]$ & $\Gamma(\chi_{b0}\rightarrow\gamma\gamma)_{LO} [keV]$ &  $\Gamma(\chi_{b0}\rightarrow\gamma\gamma)_{NLO} [keV]$\\
    \hline
    Harmonic oscillator & 4.2   & 0.053 & 0.047 & 0.048 \\
    Logarithmic         & 5.0   & 0.032 & 0.017 & 0.017\\
    Power-law           & 4.721 & 0.033 & 0.018 & 0.019 \\
    Cornell             & 5.17  & 0.028 & 0.014 & 0.014 \\
    Buchm\"ulller-Tye   & 4.87  & 0.031 & 0.017 & 0.017 \\
    \hline
    \hline
    \end{tabular}
    \label{tab:FF_00_chib}
\end{table}
\begin{table}[]
    \centering
        \caption{$|F_{TT}(0,0)|$ for $\chi_{b0}$ with quark/antiquark $m_{b}$ = 4.18 GeV according to PDG \cite{PDG}.}
    \begin{tabular}{l|c|c|c|c}
    \hline
    \hline
    Potential type  &  $|F_{TT}(0,0)|[GeV]$ & $\Gamma(\chi_{b0}\rightarrow\gamma\gamma)_{LO} [keV]$ &  $\Gamma(\chi_{b0}\rightarrow\gamma\gamma)_{NLO} [keV]$\\
    \hline
    Harmonic oscillator & 0.053 & 0.048 & 0.049 \\
    Logarithmic         & 0.045 & 0.034 & 0.035\\
    Power-law           & 0.042 & 0.030 & 0.030 \\
    Cornell             & 0.043 & 0.031 & 0.031 \\
    Buchm\"ulller-Tye   & 0.042 & 0.030 & 0.030 \\
    \hline
    \hline
    \end{tabular}
    \label{tab:FF_00_chib_mb}
\end{table}

From Eqs.~(\ref{eq:Gamma}), (\ref{eq:Gamma_R0_LO}) and (\ref{eq:Gamm_NLO}), using the radiative decay width for $\chi_{c0}$, we extracted the first derivative of radial wave function at the origin $R'(0)$, which is contained in Tab.~\ref{tab:R0_chic} and compared to result obtained from Eq. (\ref{eq:uk_to_R0}).
In Tab.~\ref{tab:R_0_chib} we listed $R'(0)$ values for $\chi_{b0}$.

In the NR limit there is an ambiguity whether calculate the decay width using the quark mass $m_Q$ or the meson mass $M_{\chi}$, as to the NR acurracy $M_{\chi} = 2 m_Q$, i.e.:
\begin{subequations}
\begin{eqnarray}
\centering
 \Gamma(\chi_{Q0} \to \gamma\gamma) &=& \frac{9\cdot2^4\alpha_{\rm em}^2
   e_{Q}^{4} N_{c}}{M^4_{\chi}} |R'(0)|^2 \; , \, \rm{or} \,  \label{eq:width_R0_Mchic}\\
 \Gamma(\chi_{Q0} \to \gamma\gamma) &=& \frac{9\alpha_{\rm em}^2 e_{Q}^{4} N_{c}}{m^4_{Q}} |R'(0)|^2 \; .    \label{eq:width_R0_mc} 
\end{eqnarray}
 \label{eq:width_R0}
\end{subequations}
We calculated $\Gamma(\chi_{Q0}\rightarrow\gamma\gamma)_{NLO}$ respectively for $\chi_{c0}$ and $\chi_{b0}$ (see Tab.~\ref{tab:R0_chic} and Tab.~\ref{tab:R_0_chib})
for five different potentials from the literature
   using first Eq.~(\ref{eq:width_R0_Mchic}) (with meson mass as in Tab.~\ref{tab:table1}) and then Eq.~(\ref{eq:width_R0_mc}) for comparison. 

\begin{table}[h!]
    \centering
        \caption{$\Gamma(\chi_{c0}\rightarrow\gamma\gamma)_{NLO}$ according to
        Eqs.~(\ref{eq:Gamm_NLO}), (\ref{eq:width_R0}) with respect of quark mass for distinguished potential. $R'(0)$ is obtained from Eq.~(\ref{eq:uk_to_R0}) and compared to $R'(0)$ extracted from experiment as explained in the text.} 
    \begin{tabular}{l|c|c|c|c}
    \hline 
    \hline
    Potential type&$m_{c}$ [GeV] & R'(0) [GeV$^{5/2}$]&$\Gamma(\chi_{c0}\rightarrow\gamma\gamma)_{NLO}$ [keV] &$\Gamma(\chi_{c0}\rightarrow\gamma\gamma)_{NLO}$[keV] \\
    & & & Eq.~(\ref{eq:width_R0_Mchic})& Eq.~(\ref{eq:width_R0_mc})\\
    \hline
     Harmonic oscillator & 1.4  & 0.27 & 2.42 & 5.54 \\
     Logarithmic         & 1.5  & 0.24 & 1.85 & 3.11 \\
     Powerlike           &1.334 & 0.22 & 1.62 & 4.34 \\
     Cornell             & 1.84 & 0.32 & 2.51 & 3.38 \\
     Buchm\"uller-Tye    & 1.48 & 0.25 & 2.15 & 3.81 \\
\hline
     extracted from experiment \cite{PDG} & &$0.25\pm 0.01$ &$2.20\pm 0.16$ &  \\
     \hline
     \hline
    \end{tabular}
    \label{tab:R0_chic}
\end{table}

\begin{table}[]
    \centering
    \caption{$\Gamma(\chi_{b0}\rightarrow\gamma\gamma)_{NLO}$ according to
        Eqs.~(\ref{eq:Gamm_NLO}), (\ref{eq:width_R0}) with respect of quark mass for distinguished potential. $R'(0)$ is obtained from Eq.~(\ref{eq:uk_to_R0}) and compared to $R'(0)$ extracted from experiment as explained in the text.}
    \begin{tabular}{l|c|c|c|c} 
    \hline
    \hline
     Potential type&$m_{b}$ [GeV] & R'(0) [GeV$^{5/2}$] & $\Gamma(\chi_{b0}\rightarrow\gamma\gamma)_{NLO} [keV]$ &  $\Gamma(\chi_{b0}\rightarrow\gamma\gamma)_{NLO} [keV]$ \\
         & & & Eq. (\ref{eq:width_R0_Mchic})& Eq. (\ref{eq:width_R0_mc})\\
     \hline
     Harmonic oscillator            & 4.2   & 1.07 & 0.035 & 0.066\\
     Logarithmic                    & 5.0   & 1.22 & 0.045 & 0.043\\
     Powerlike                      & 4.721 & 0.98 & 0.029 & 0.035\\
     Cornell                        & 5.17  & 1.37 & 0.057 & 0.047\\
     Buchm\"uller-Tye               & 4.87  & 1.13 & 0.038 & 0.041\\
    \hline
    \hline
    \end{tabular}
    \label{tab:R_0_chib}
\end{table}

\section{The $\gamma^* \gamma^* \to \chi_{c0}$ and $\gamma^* \gamma^* \to \chi_{b0}$ form factors: numerical results}
\label{sec:EM_FF_plots}

Let us now focus for a while on the form factors calculated with the help
of the light-cone wave functions (see Eqs.~(\ref{eq:G1_G2}), (\ref{eq:FTT_FLL})).

\begin{figure}[h!]
    \centering
    \includegraphics[width=0.3\linewidth]{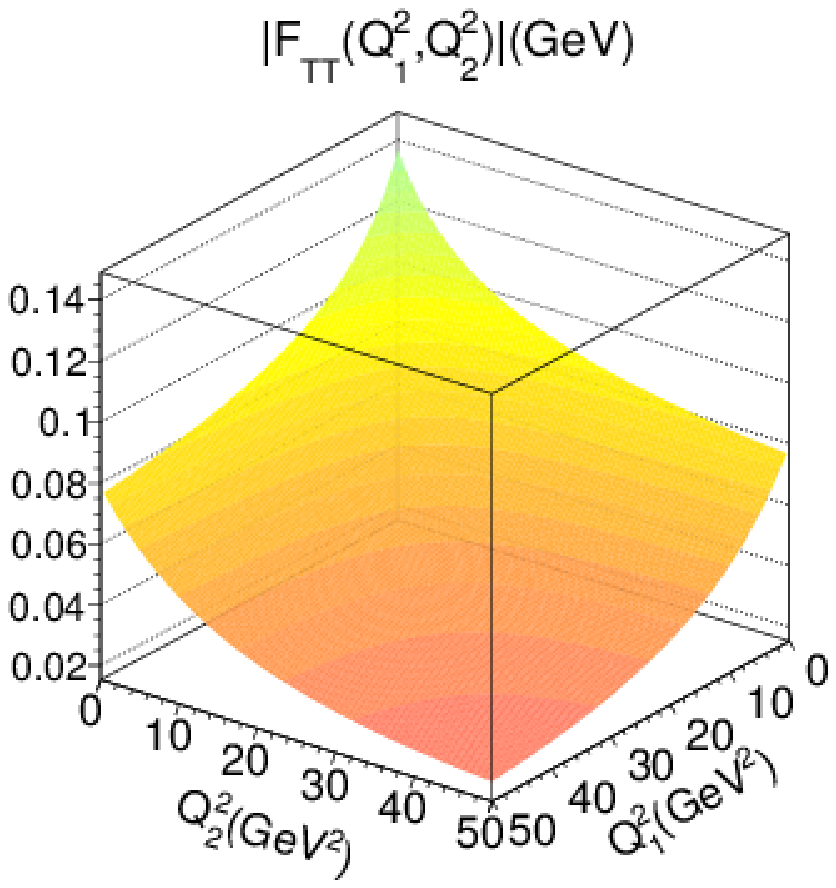}
    \includegraphics[width=0.3\linewidth]{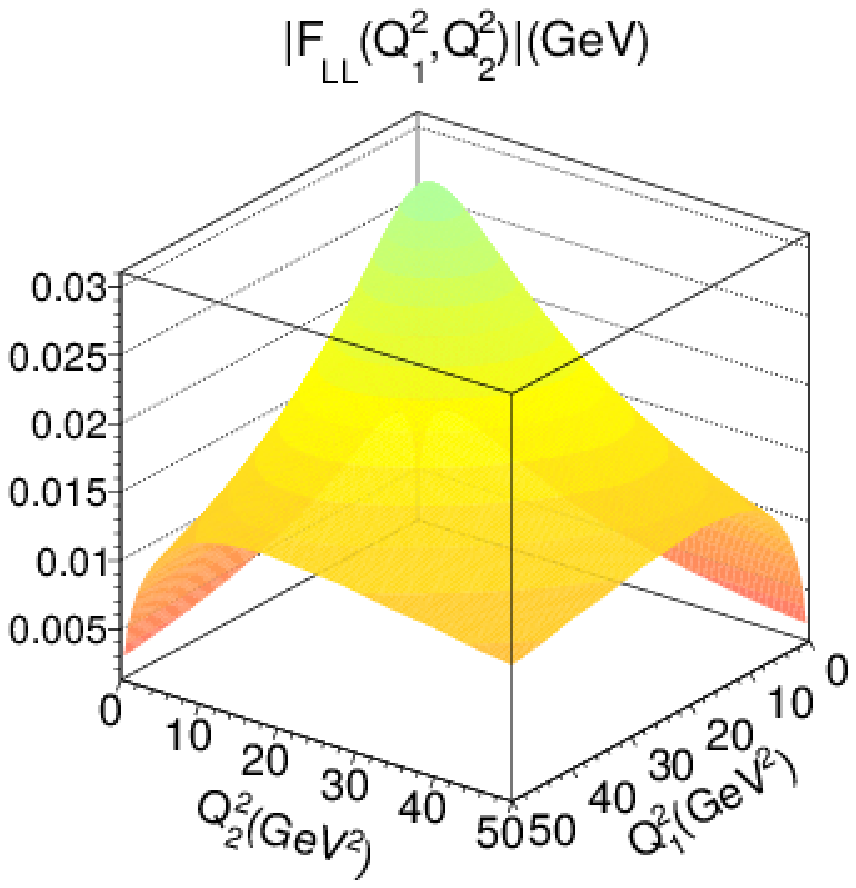}
    \includegraphics[width=0.3\linewidth]{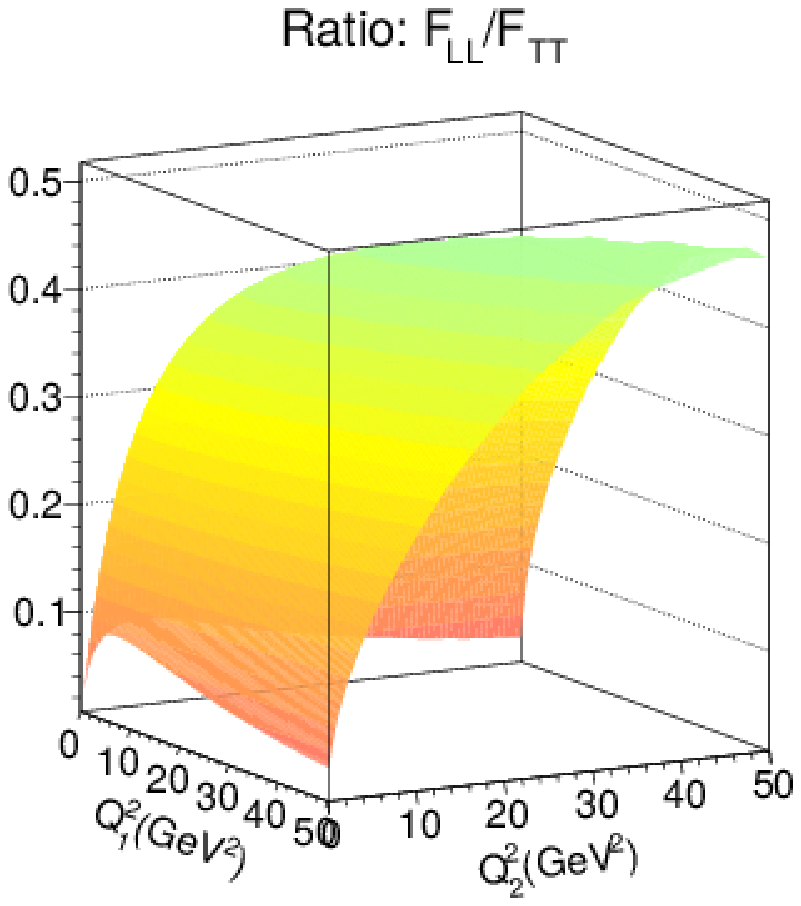}
    \caption{$|F_{TT}|$ (left) and $F_{LL}$ (middle) form factors
     and their ratio (right) for the Buchm\"uller-Tye potential for 
    $\chi_{c0}$.}
    \label{fig:F_TT_and_F_LL}
\end{figure}

\begin{figure}
    \centering
    \includegraphics[width= 0.3\linewidth]{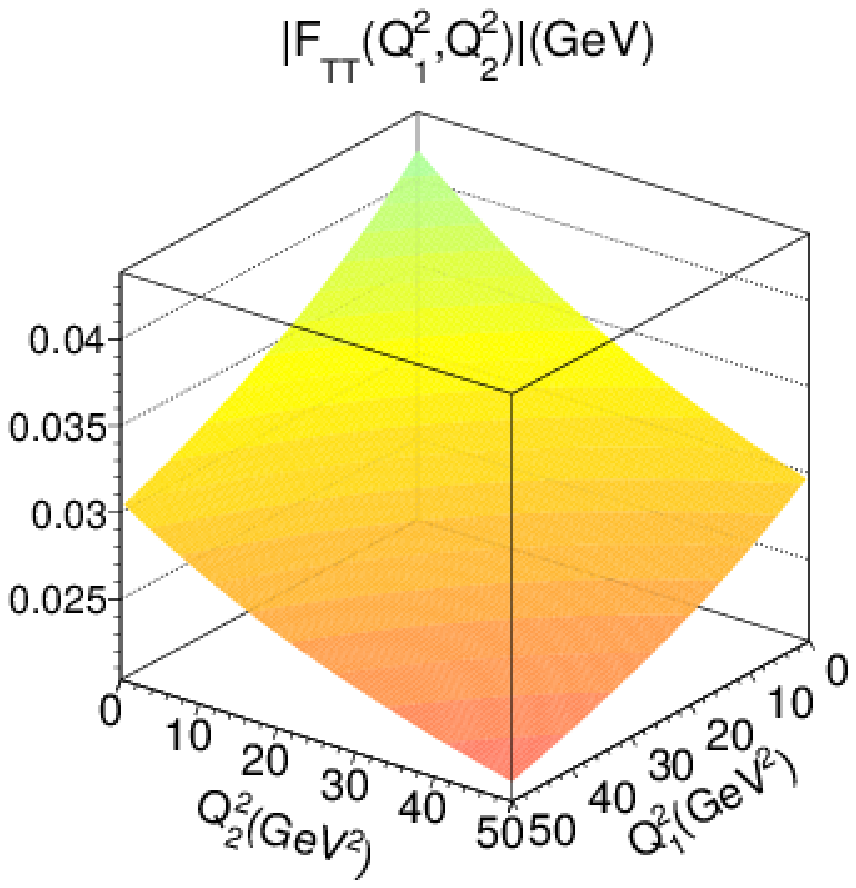}
    \includegraphics[width= 0.3\linewidth]{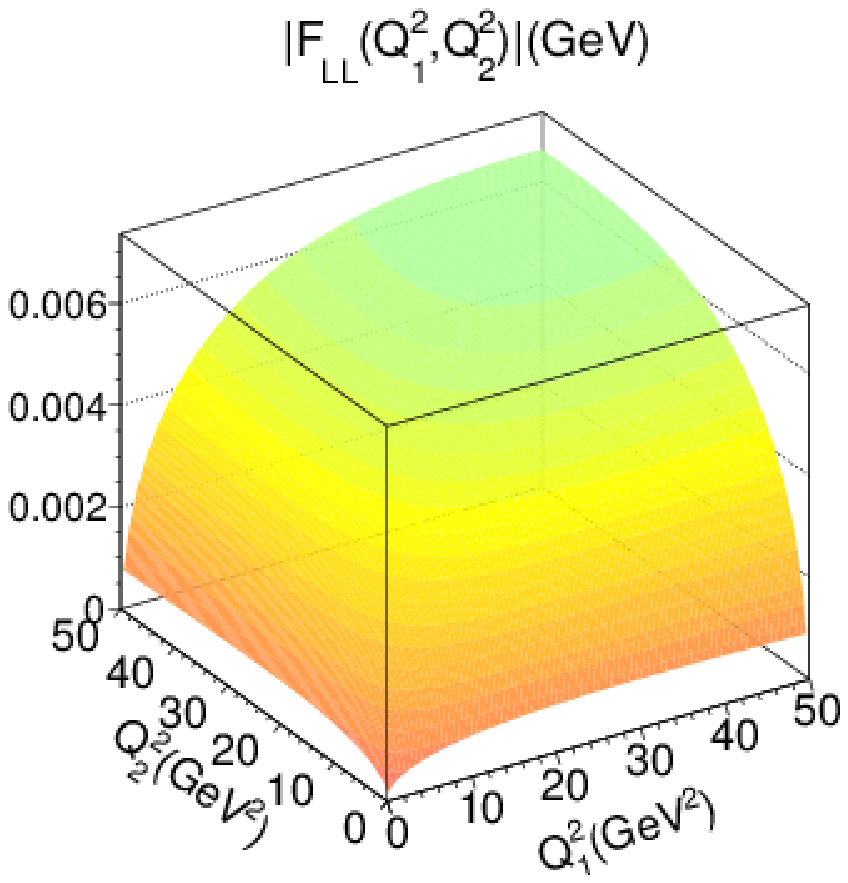}
    \includegraphics[width = 0.3\linewidth]{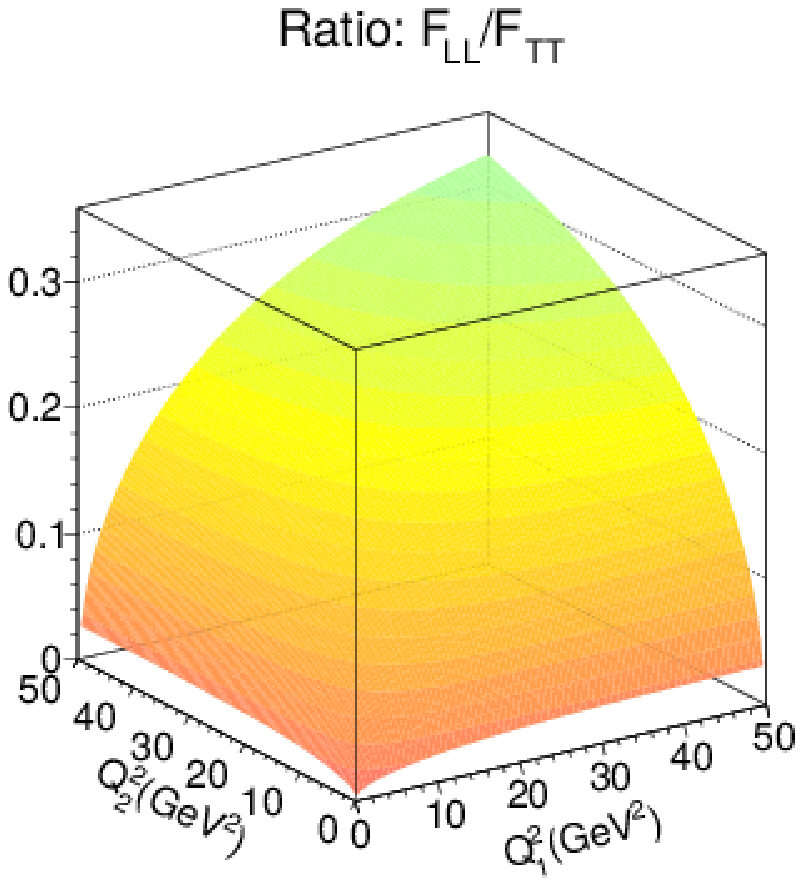}
    \caption{$|F_{TT}|$ (left) and $F_{LL}$ (right) form factors
     for the Buchm\"uller-Tye potential for $\chi_{b0}$.}
    \label{fig:F_TT_and_F_LL_chib}
\end{figure}

The form factors $F_{TT}$ and $F_{LL}$, which have a more straightforward interpretation than $G_1,G_2$, are
shown in Fig.~\ref{fig:F_TT_and_F_LL} for $\chi_{c0}$ and 
Fig.~\ref{fig:F_TT_and_F_LL_chib} for $\chi_{b0}$. There is a smooth 
dependence on virtualities with maxima at $Q_1^2 = 0$ and $Q_2^2 = 0$
for both mesons.
One can see the required Bose symmetry under exchange of $Q_1^2$ and $Q_2^2$.
The two form factors have a quite different
dependence on photon virtualities. This is demonstrated in the right
panel of Fig.~\ref{fig:F_TT_and_F_LL} where the $F_{LL}$-to-$F_{TT}$
ratio is shown.
We note, that $F_{LL} \to 0$ when either~$Q_1^2 \to 0$~or~$Q_2^2 \to 0$.

\begin{figure}[h!]
    \centering
    \includegraphics[width=0.3\linewidth]{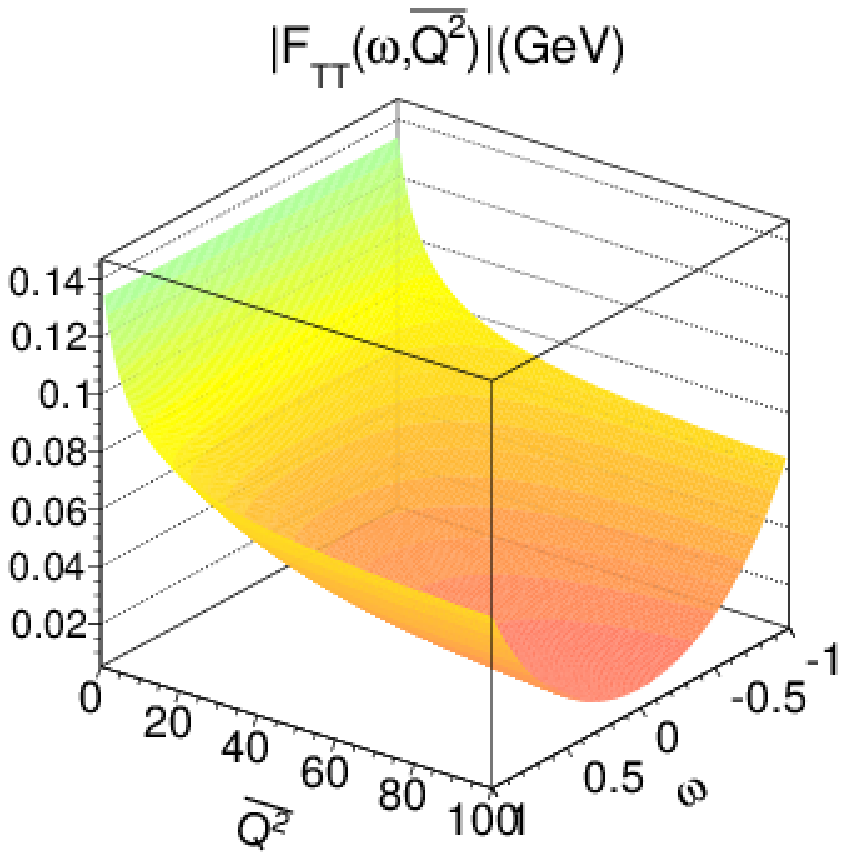}
    \includegraphics[width=0.3\linewidth]{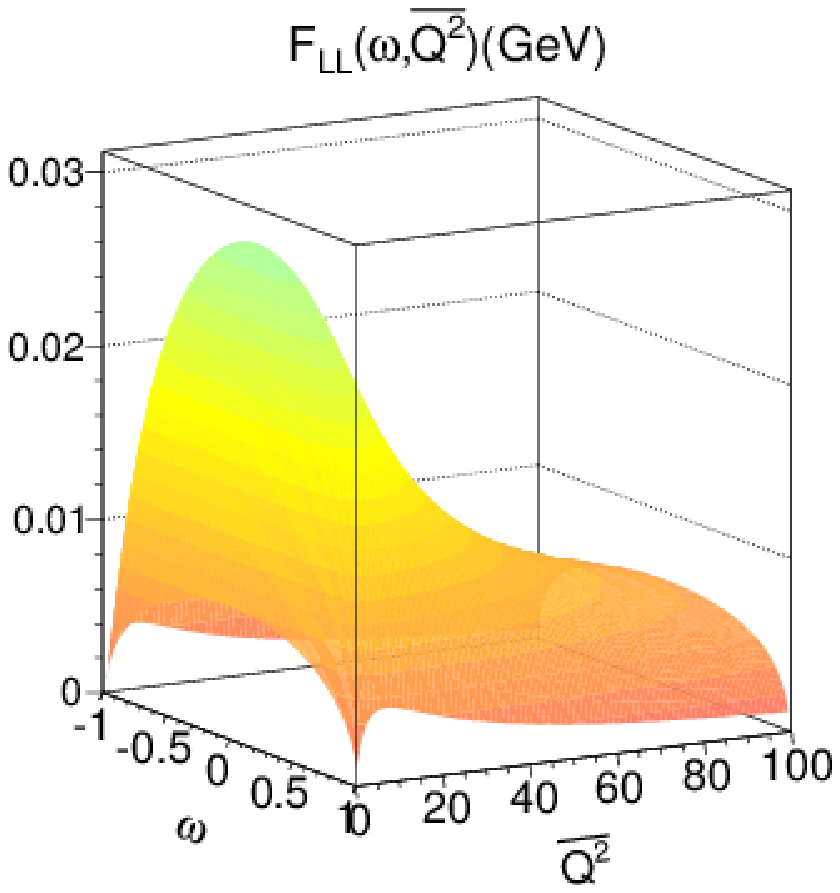}
    \caption{$|F_{TT}|$ (left) and $F_{LL}$ (right) form factors
     for the Buchm\"uller-Tye potential as a function of $\omega$ 
     and $\bar{Q^2}$ for $\chi_{c0}$.}
    \label{fig:F_TT_and_F_LL_Q2omega}
\end{figure}

In order to compare to the situation for $\gamma^* \gamma^* \to
\eta_c(1S,2S)$ discussed in Ref.~\cite{BGPSS2019}, in Fig.~\ref{fig:F_TT_and_F_LL_Q2omega} 
we show the two form factors as a function of 
\begin{eqnarray}
 \overline{Q}^2 = {Q_1^2 + Q_2^2 \over 2} \,  \, {\rm{and}} \,  \, \omega = \frac{Q_1^2 - Q_2^2}{Q_1^2 + Q_2^2} \, .
\end{eqnarray}
In contrast to the case of $\eta_c$ production \cite{BGPSS2019} we do
not observe the scaling in $\omega$, neither for $F_{TT}$, nor $F_{LL}$.

\begin{figure}
    \centering
    \includegraphics[width = 0.45\linewidth]{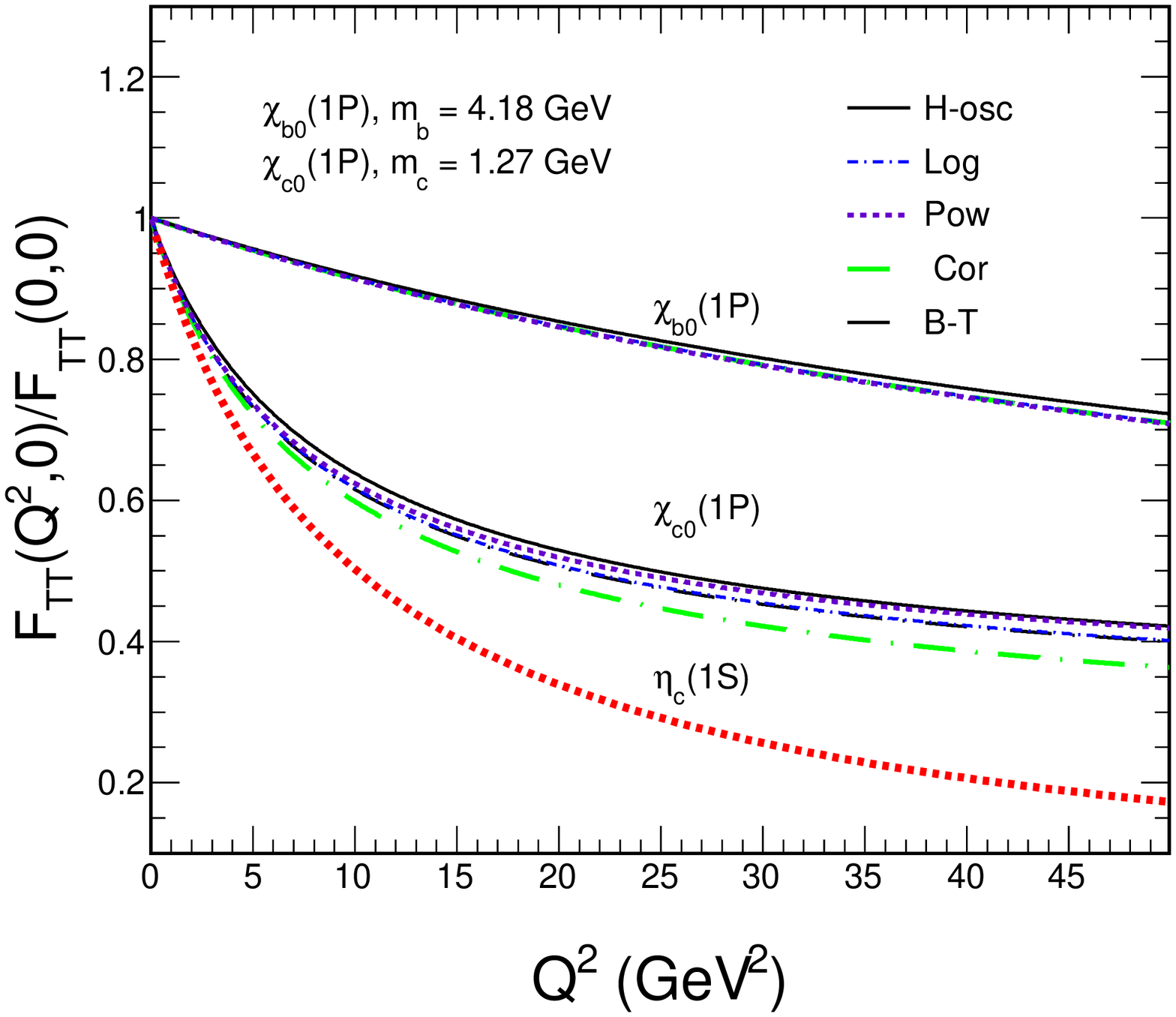}
    \includegraphics[width = 0.45\linewidth]{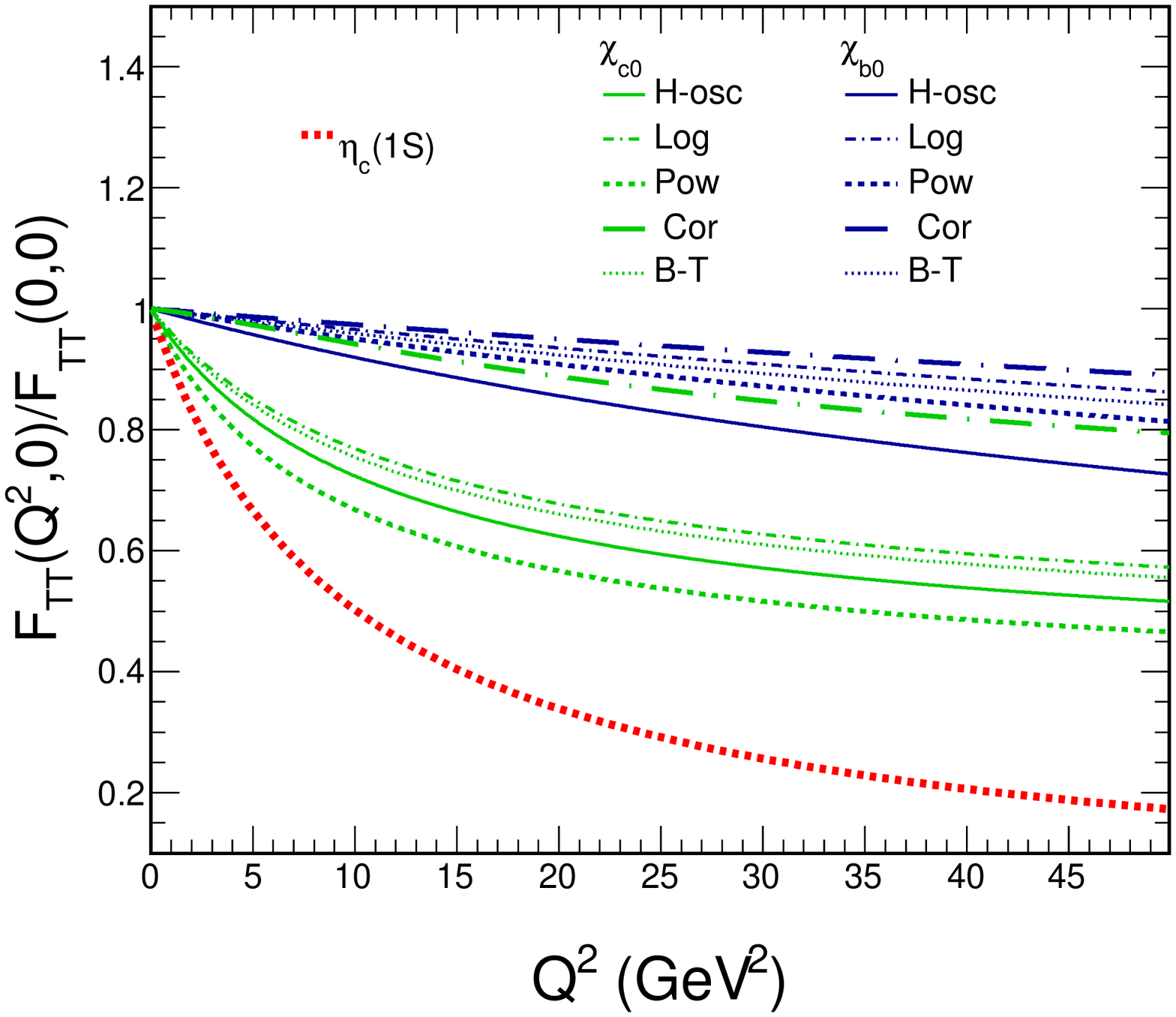}
    \caption{
  Normalized to unity TT form factor for three different mesons: 
             $\eta_{c}$(1S), $\chi_{c0}$(1P), $\chi_{b0}$(1P).}
    \label{fig:FF_F0}
\end{figure}

In the left panel of Fig.~\ref{fig:FF_F0} we compare the normalized form
factors $F_{TT}(Q^2,0)/F_{TT}(0,0)$ for 
$\eta_c(1S)$ with its counterpart for $\chi_{c0}(1P)$ and $\chi_{b0}(1P)$ 
in the virtuality interval $Q^2 \in$ (0,50 GeV$^2$). For the form factor of the $\chi_{c0}$ 
we used the mass of the $c$-quark $m_c = 1.27$ GeV, and for the $\chi_{b0}$ form factor, $m_{b}= 4.18$ GeV. Results for five different potential
models are presented for $\chi_{c0}(1P)$ and $\chi_{b0}(1P)$, whereas for the case of 
$\eta_c$(1S), we show the result from Ref.~\cite{BGPSS2019} for the power-law potential model only. 
In the right panel we present the normalized form factors for five different
potential models for $\chi_{c0}$ and $\chi_{b0}$, but with corresponding quark masses specific to each potential model.

A first measurement in $e^+ e^-$ collisions of $\chi_{c0}$ production and single-tag
$Q^2$ dependence of the cross section was done recently by the Belle collaboration \cite{Masuda:2017rhm}
in the $K_S K_S$ final state.
The statistics is, however, too low (about 10 events) to allow  detailed studies.

\section{Differential cross sections for the $p p \to \chi_{c0}$ and  $p p \to \chi_{b0}$ reactions}
\label{sec:hadroproduction}

In this subsection we shall present differential distributions in
the $\chi_{c0}$ transverse momentum and rapidity for the $p p \to \chi_{c0}$ 
reaction as well as distributions in the same observables for $p p \to \chi_{b0}$.
Here we shall set $\sqrt{s}$ = 13 TeV.
The grids for $G_1$ and $G_2$ form factors are used in an interpolation procedure to obtain values relevant for the cross
section calculations.

\begin{figure}
    \centering
    \includegraphics[width = 0.45\linewidth]{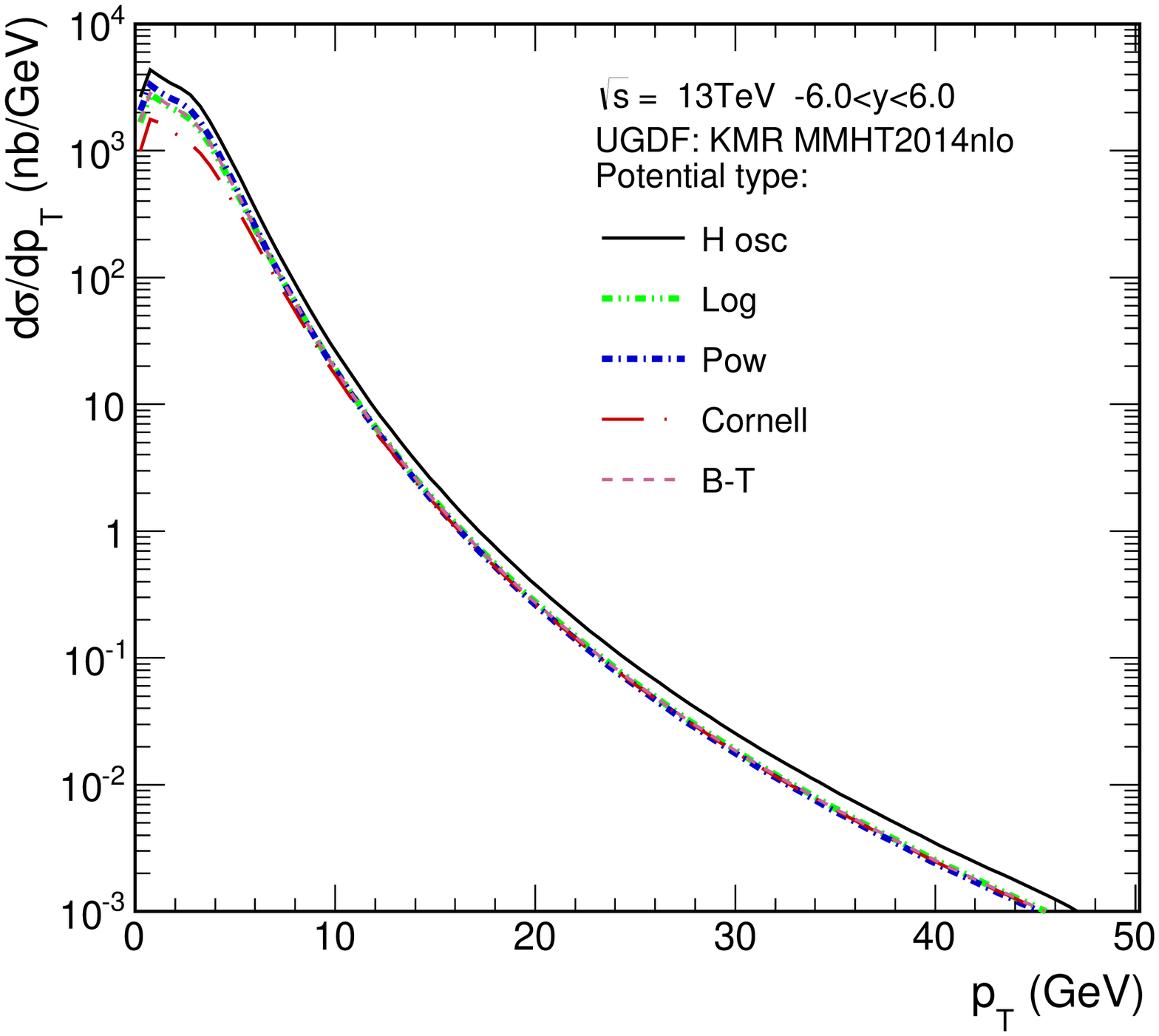} 
    \includegraphics[width = 0.45\linewidth]{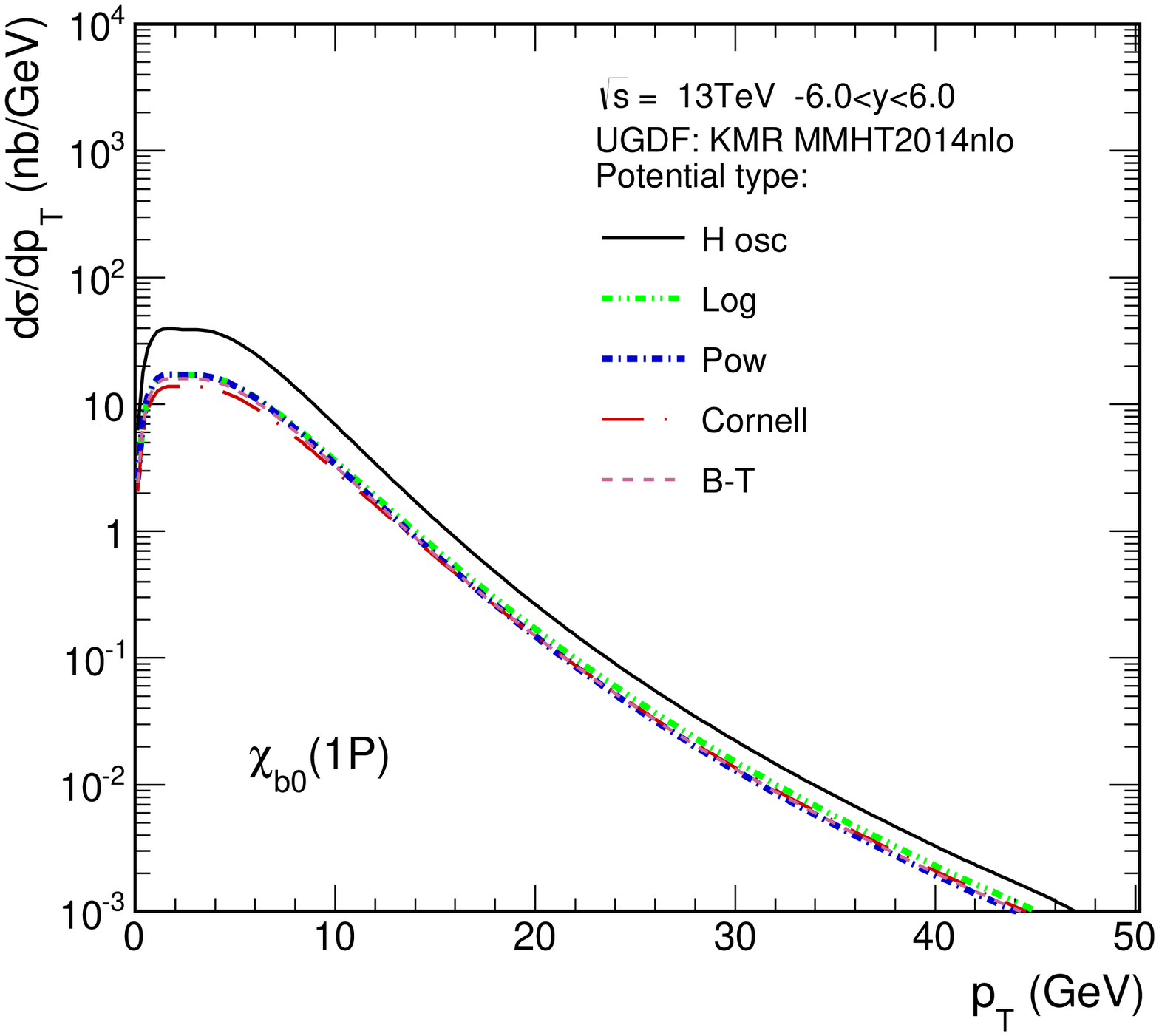}\\
    \includegraphics[width = 0.45\linewidth]{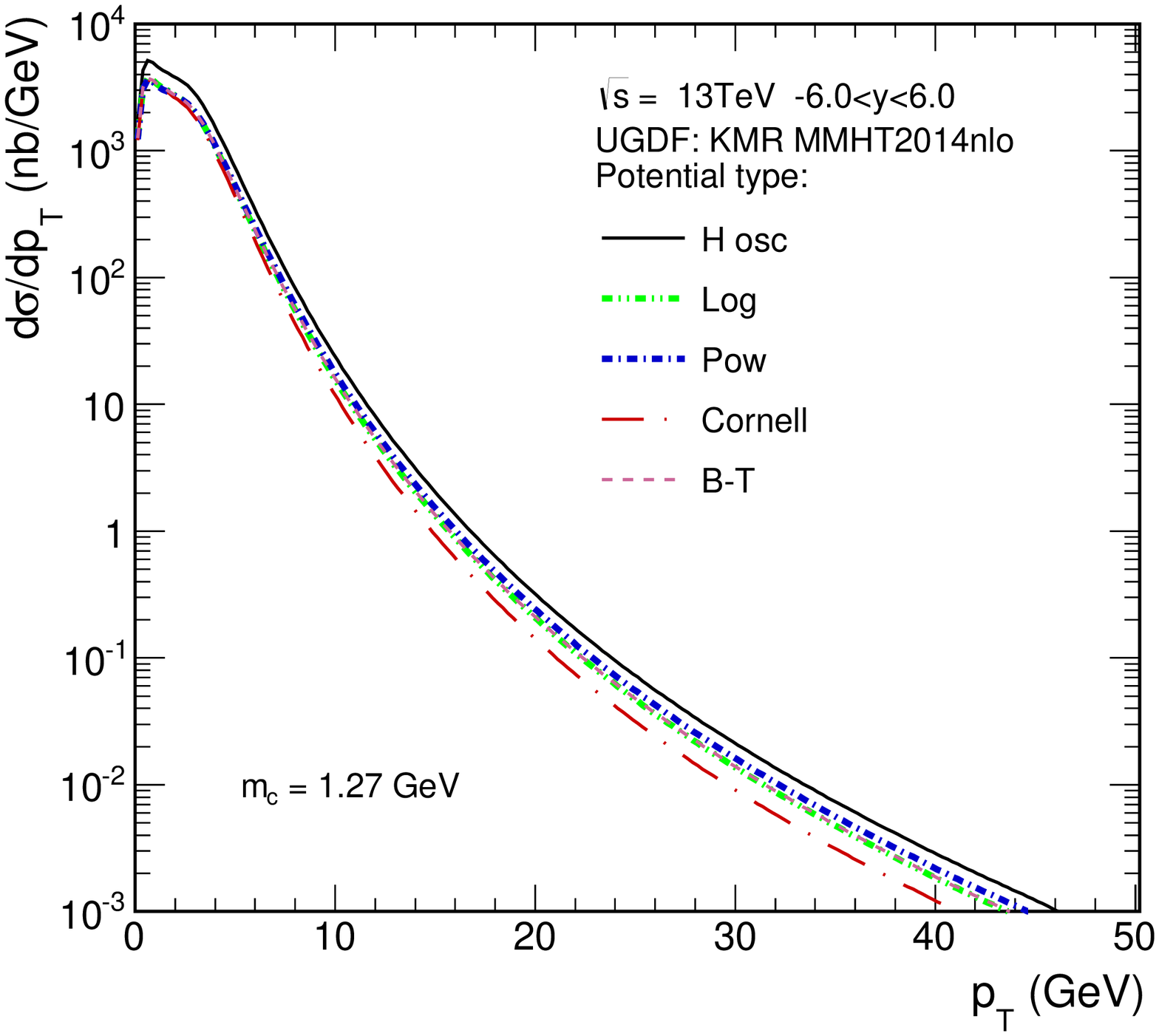}
    \includegraphics[width = 0.45\linewidth]{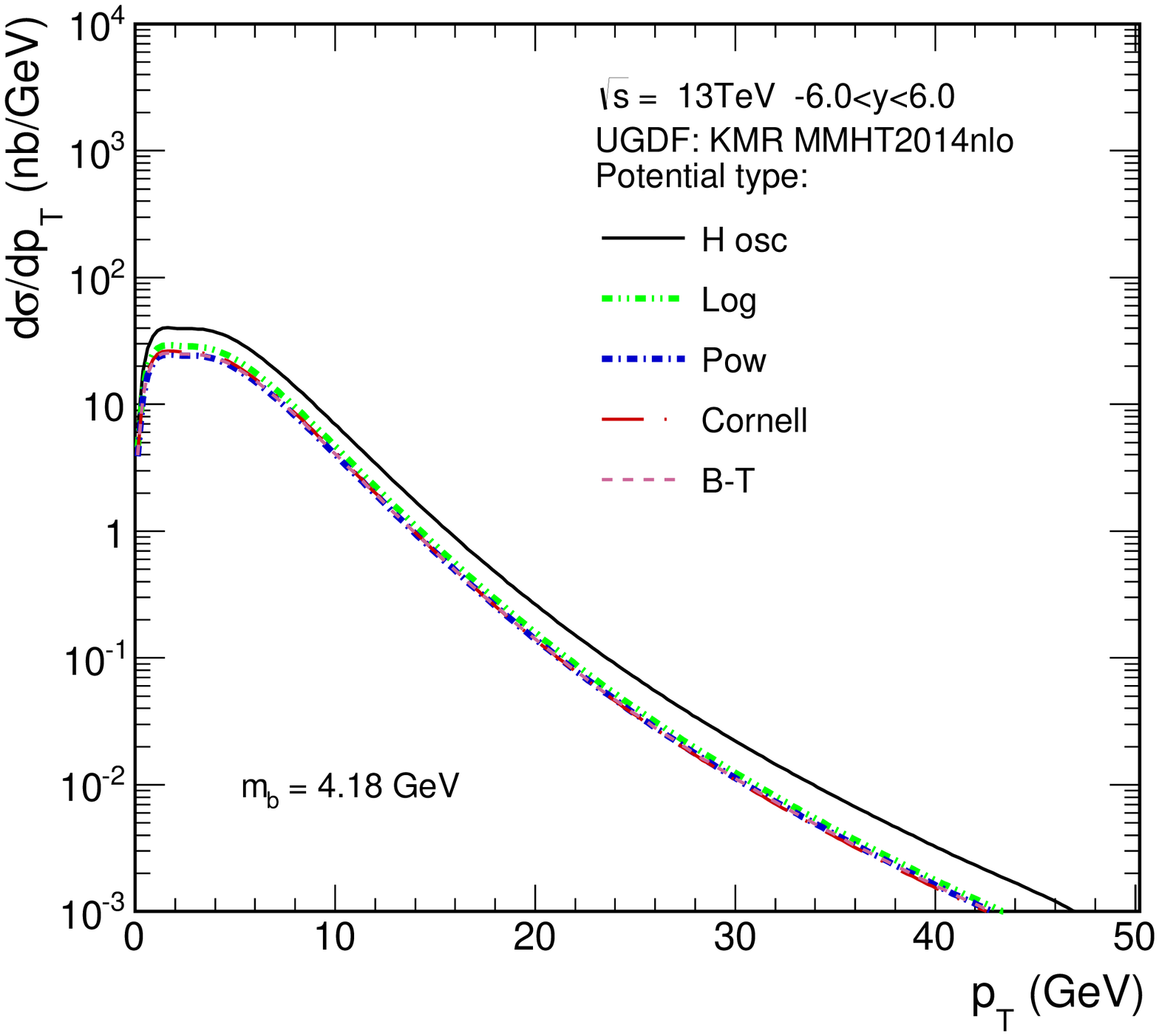}
    \caption{Comparison of $p_{T}$ distributions for five different
      potentials types with corresponding quark mass in the first row, 
      and the second row is with identical (for all potentials) 
      $m_{c} = 1.27$ GeV and $m_{b} = 4.18$ GeV used in calculating 
      form factors.
      Here the KMR  UGDF was used for $\chi_{c0}$(1P) (left panel) and 
      $\chi_{b0}$(1P) (right panel).}
    \label{fig:dsig_dpt_all}
\end{figure}

In Fig.\ref{fig:dsig_dpt_all} we show transverse momentum distributions
of $\chi_{c0}$ and $\chi_{b0}$ for the broad range of its rapidity 
$-6 < y < 6$ for different $c {\bar c}$ and $b {\bar b}$ potentials  specified in the figure. 
There is a rather small dependence on the potential used, except for the
region of $p_{T} \lesssim 5 \, \rm{GeV}$.
In the first row one can find the plots where we used in the form factor the quark mass
corresponding to the respective $q{\bar q}$ potential model.
The second row presents the plots with 
$m_{c} = 1.27 \, \rm{GeV}$ and $m_{b} = 4.18 $ GeV. 
In this case, in the region of meson 
$p_{T} <$ 5 GeV there is no strong dependence on the $q{\bar q}$ 
potential model, but in the region of $p_T > 5 \, \rm{GeV}$ for the $\chi_{c0}$ one can observe a spread of the curves.

\begin{figure}
    \centering
    \includegraphics[width = 0.45\linewidth]{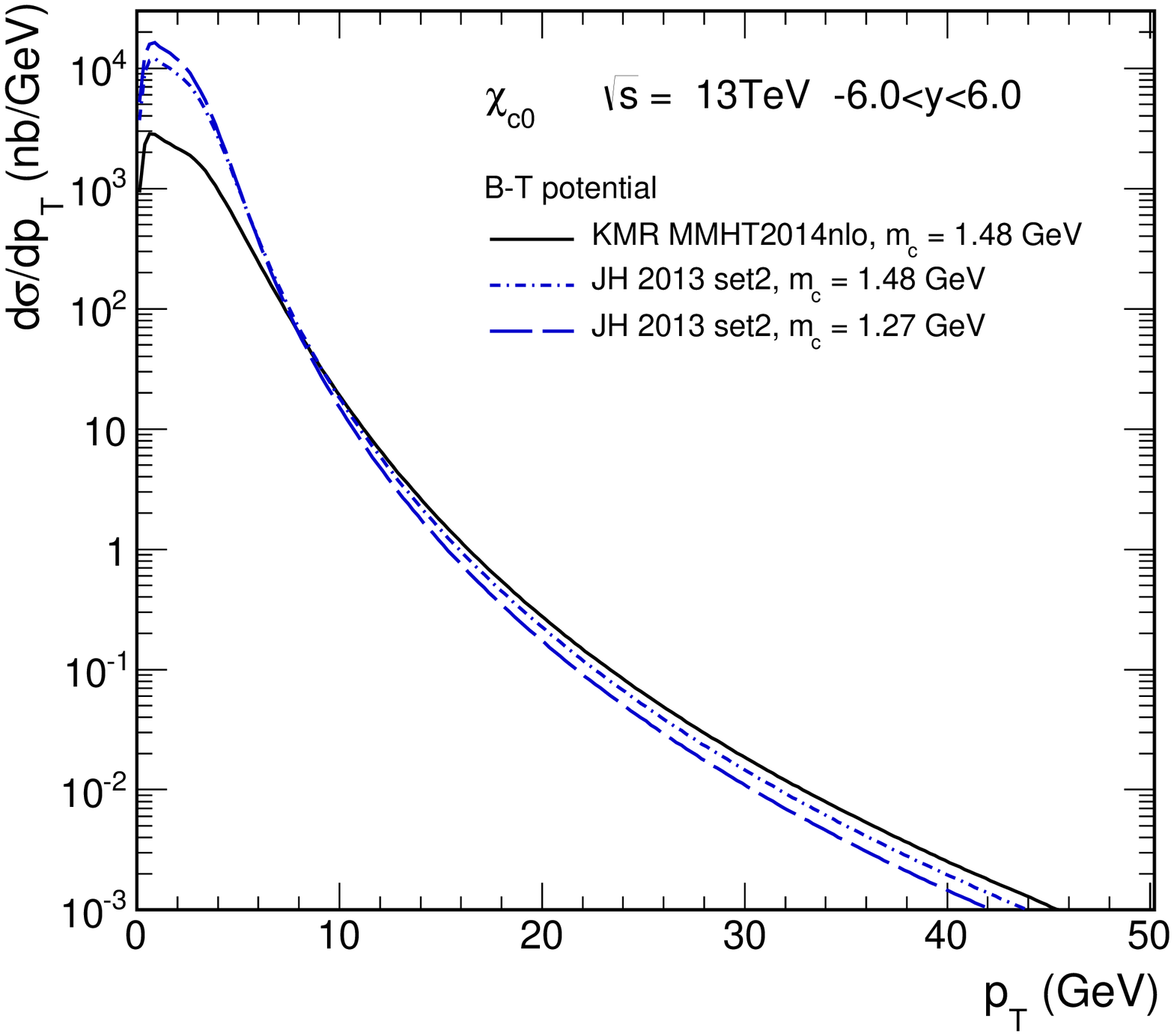}
    \includegraphics[width = 0.45\linewidth]{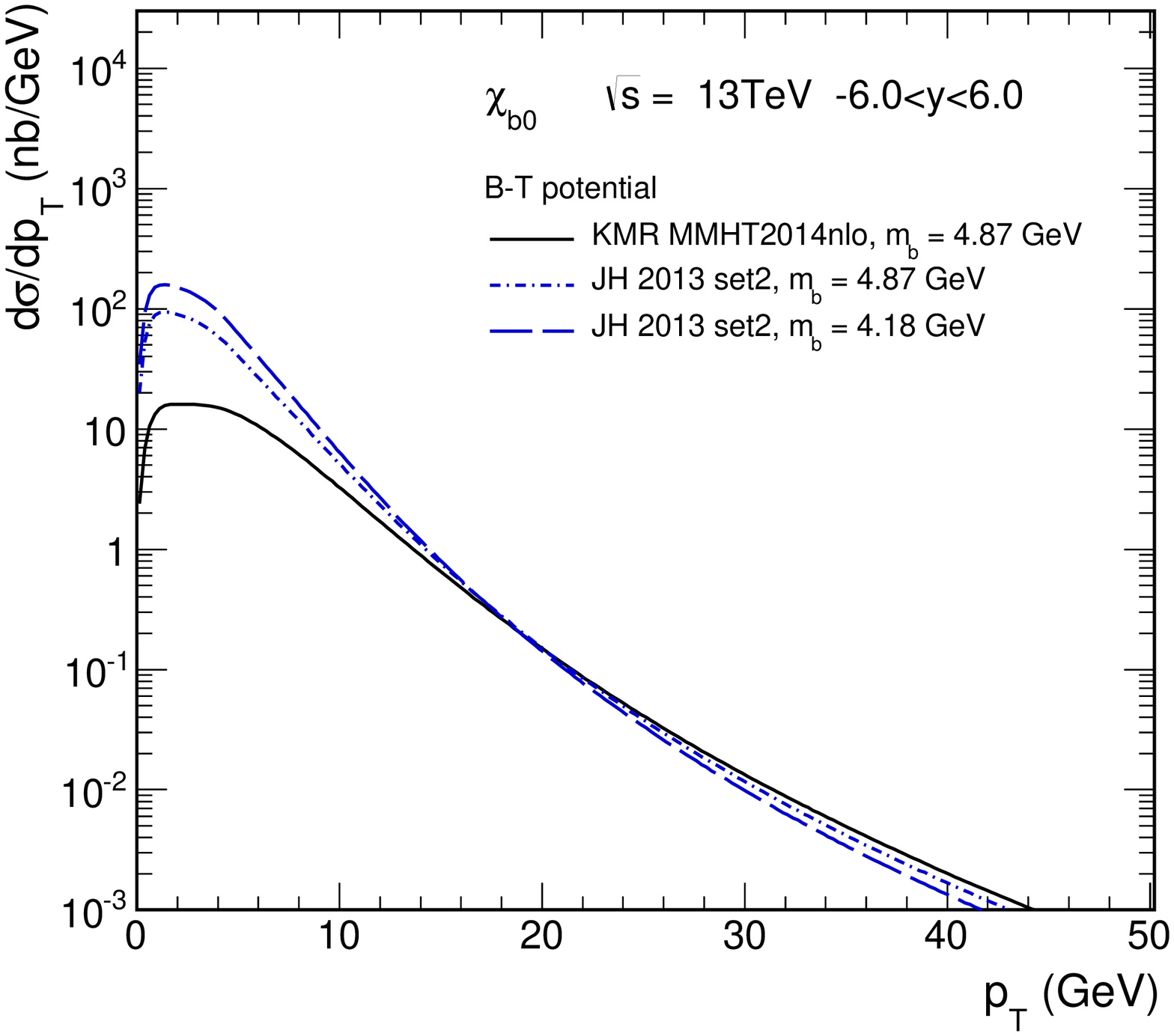}
    \caption{Comparison of $p_{T}$ distributions of $\chi_{c0}$ (left)
    and $\chi_{b0}$ (right) for two different UGDFs.
      Here the Buchm\"uller-Tye potential was used.}
    \label{fig:dsig_dpt_UGDF}
\end{figure}

In Fig.\ref{fig:dsig_dpt_UGDF} we compare results obtained with two different
UGDs.
The  UGDs, which we labeled in figures as "KMR"
are based on the prescription \cite{Kimber:2001sc} with updates \cite{Martin:2009ii}.
The KMR prescription requires as an input a collinear gluon distribution.
For this purpose, we used the NLO collinear gluon distribution from \cite{Harland-Lang:2014zoa}.
We also employed the UGD "JH-2013-set2" constructed 
from HERA data on deep inelastic structure functions as a solution of the CCFM evolution equation
\cite{CCFM},
which has been obtained by Hautmann and Jung \cite{Hautmann:2013tba}.
The corresponding results differ at small transverse
momenta $p_T$, in the peak region of the distribution.  A similar behaviour was observed recently for the production of $\eta_c$ in \cite{BPSS2019}.

\begin{figure}
    \centering
    \includegraphics[width = 0.45\linewidth]{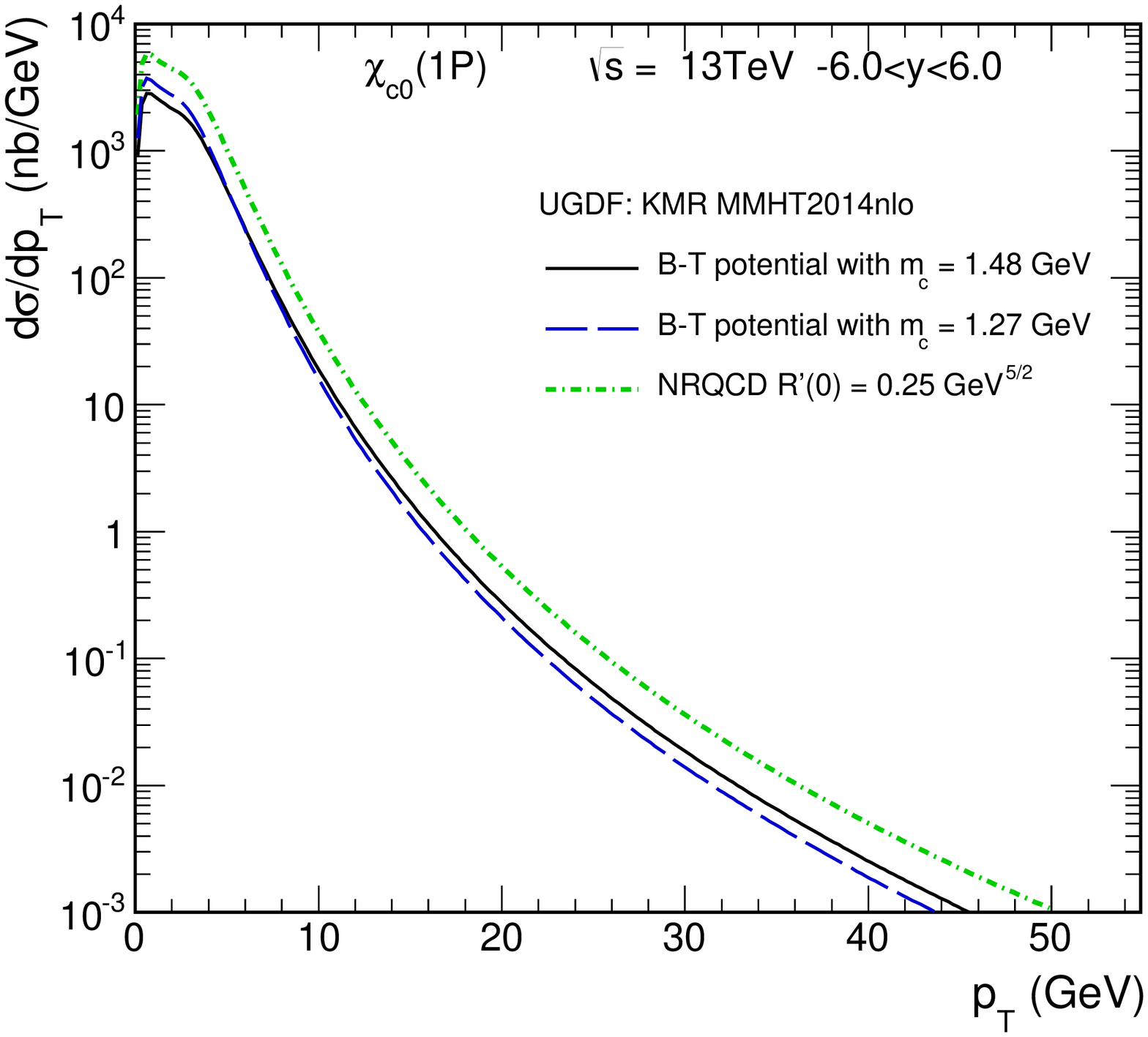}
        \includegraphics[width = 0.45\linewidth]{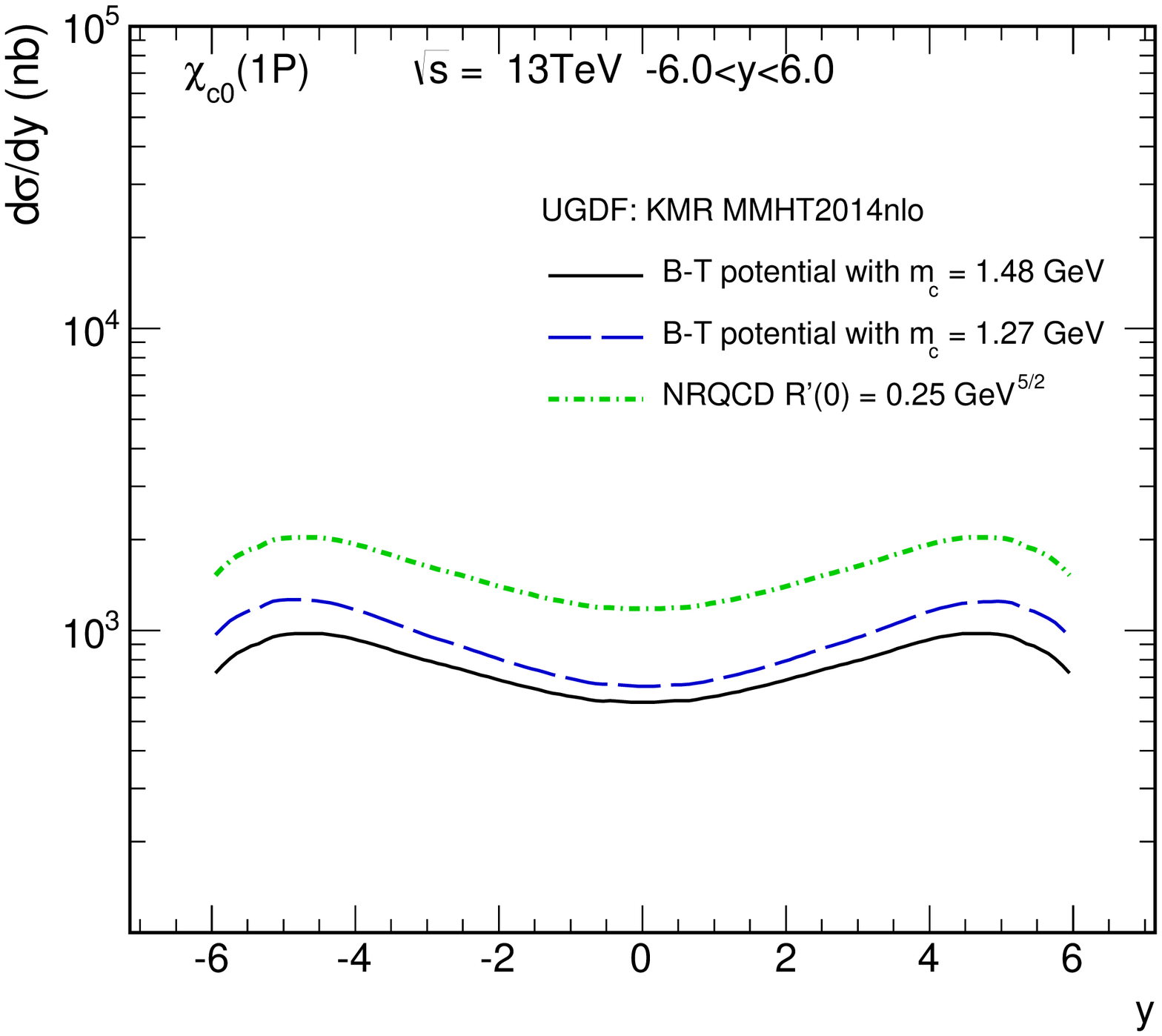}
    \caption{Comparison of $p_{T}$ distributions of $\chi_{c0}$ for 
      our full model (solid line - with $m_c = 1.48$ GeV, 
      dashed line - with $m_c = 1.27$ GeV) with the NRQCD result 
      (see Eq.(\ref{eq:tau_NRQCD})) (dash-dotted line). 
      Here the B-T potential and the KMR UGDF were used. In the NRQCD 
      matrix element we applied $R'(0) = 0.25 \, {\rm GeV}^{5/2}$ (the extracted value 
      from the experimental radiative decay width, see Tab. \ref{tab:R0_chic}).}
    \label{fig:dsig_dpt_NRQCD_chic}
\end{figure}

\begin{figure}
    \centering
    \includegraphics[width = 0.45\linewidth]{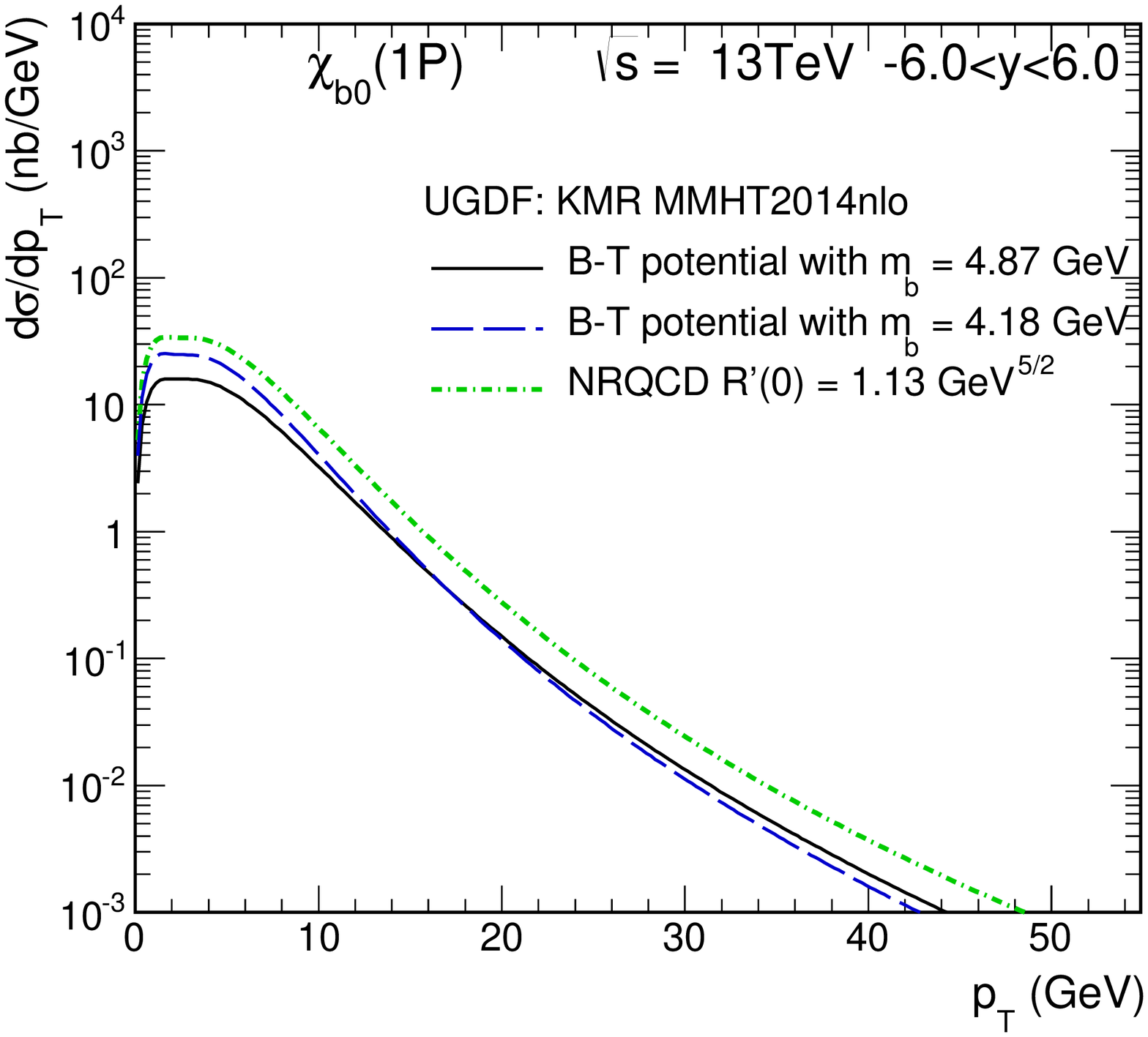}
        \includegraphics[width = 0.45\linewidth]{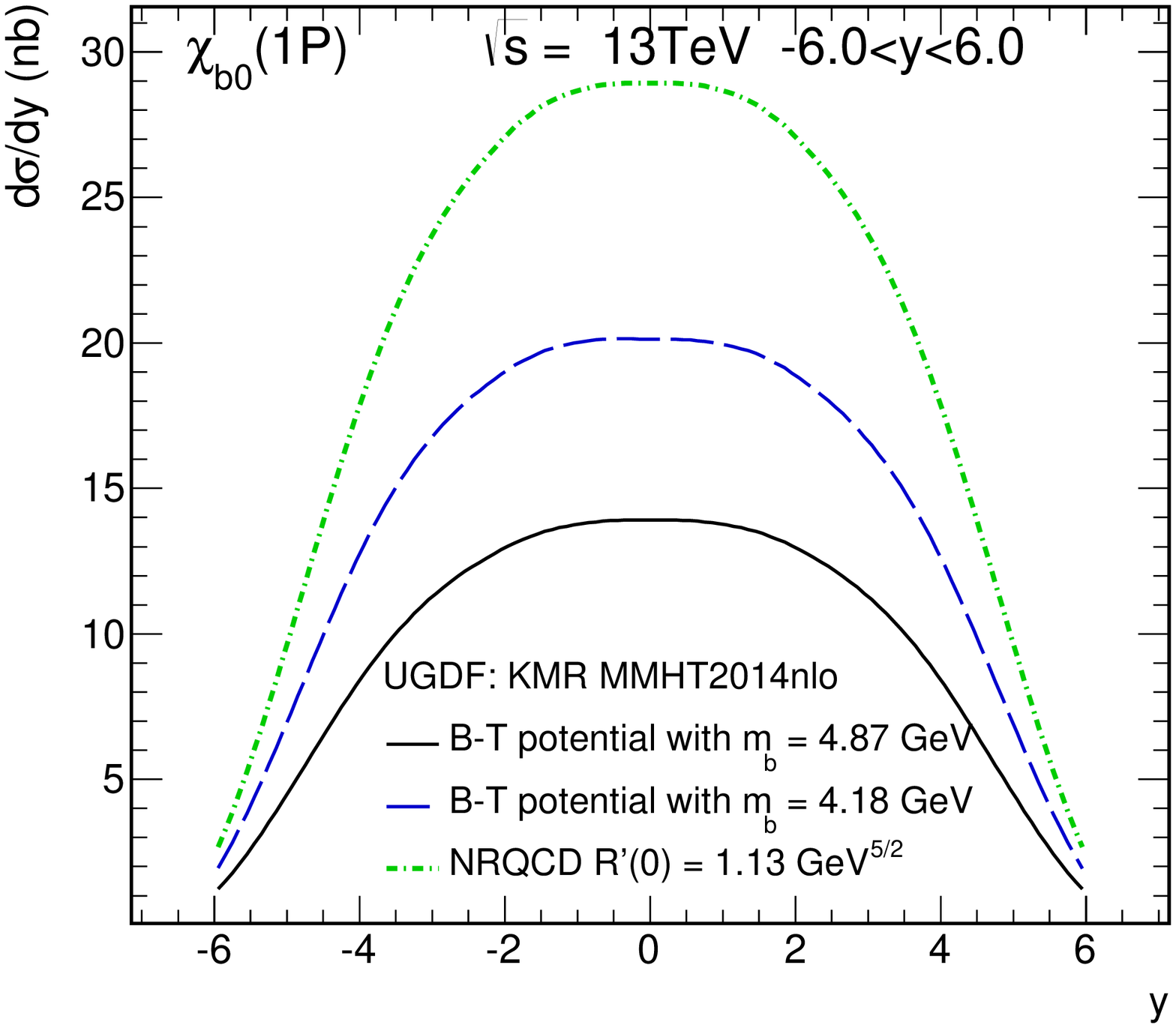}
    \caption{Comparison of $p_{T}$ distributions for our full model
      (solid line - with $m_b = 4.87$ GeV, dashed line - with $m_b = 4.18$ GeV)
      with the NRQCD result (see Eq.(\ref{eq:tau_NRQCD})) (dash-dotted line).
      Here the BT potential and the KMR  UGDF are used.}
    \label{fig:dsig_dpt_NRQCD_chib}
\end{figure}

In the present analysis we explicitly calculated the 
$\gamma^* \gamma^* \to \chi_{c0}$ as well as $\gamma^* \gamma^* \to \chi_{b0}$
form factors as presented in the previous section. Here we wish to
compare our results of corresponding cross sections with the result 
obtained previously \cite{Kniehl:2006sk} within the NRQCD approximation. 
In Fig.~\ref{fig:dsig_dpt_NRQCD_chic} and Fig.~\ref{fig:dsig_dpt_NRQCD_chib} 
we compare our full result obtained with the Buchm\"uller-Tye potential 
with that of the NRQCD approach. Both results were obtained with the KMR UGDF.

\begin{figure}
    \centering
    \includegraphics[width=0.45\linewidth]{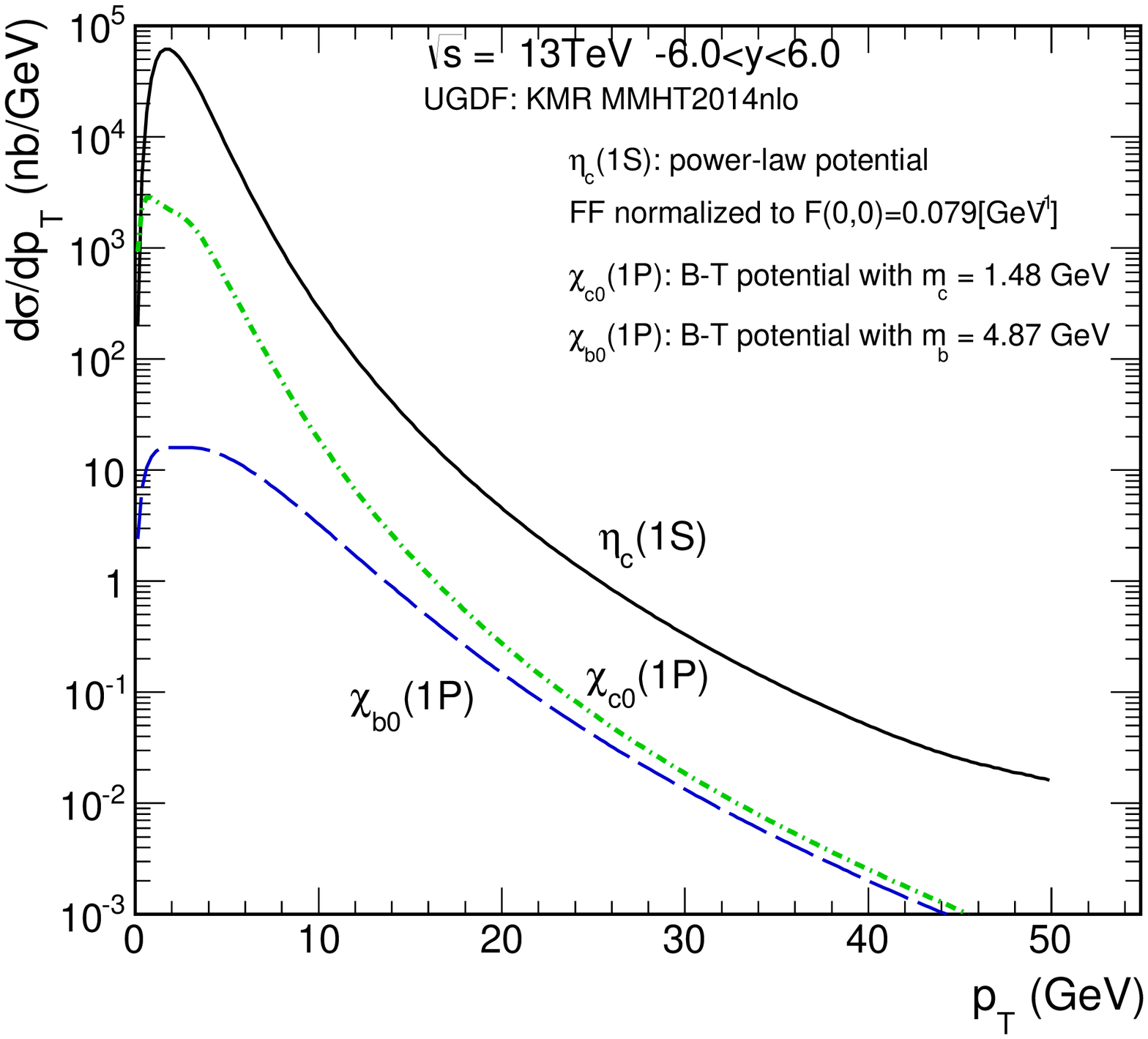}
    \includegraphics[width=0.45\linewidth]{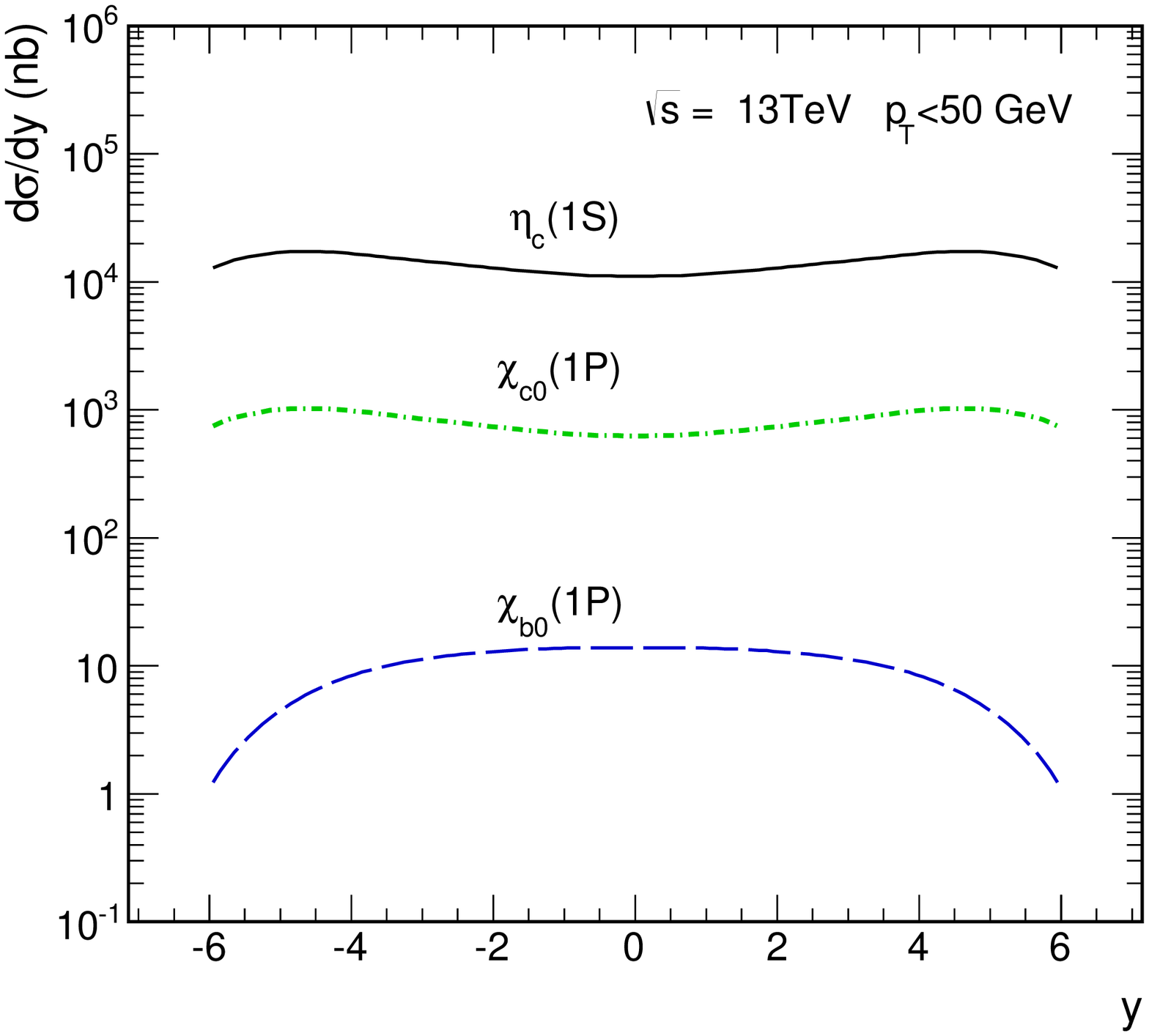}
    \caption{Differential cross section in terms of transverse momentum (lhs) and rapidity (rhs)  of the meson ($\eta_{c}$, $\chi_{c0}$, 
      $\chi_{b0}$). }
    \label{fig:meson_comparison}
\end{figure}

In Fig.\ref{fig:meson_comparison} we compare our results for $\chi_{c0}$
and $\chi_{b0}$ production with that for $\eta_c(1S)$ production from our 
recent analysis \cite{BPSS2019}. The cross section for $\chi_{c0}$
production is almost an order of magnitude smaller than that for
$\eta_c(1S)$ production. While the distributions in rapidity have a similar
shape, the distributions in transverse momentum are somewhat different.
The different shape of transverse momentum distribution is a reflection
of different matrix elements in both cases.

Since in addition, for the branching ratios to the $p \bar p$ channel, we have $Br(\chi_{c0} \to p {\bar p}) < Br(\eta_c \to p {\bar p}$), 
it may be very difficult to observe the $\chi_{c0}$ quarkonium in 
the proton-antiproton final state.
Here the $\gamma \gamma$ final state could be a better option. 
The feasibility of such a measurement requires further Monte Carlo studies 
which go beyond the scope of the present paper.

\begin{figure}
    \centering
    \includegraphics[width=0.45\linewidth]{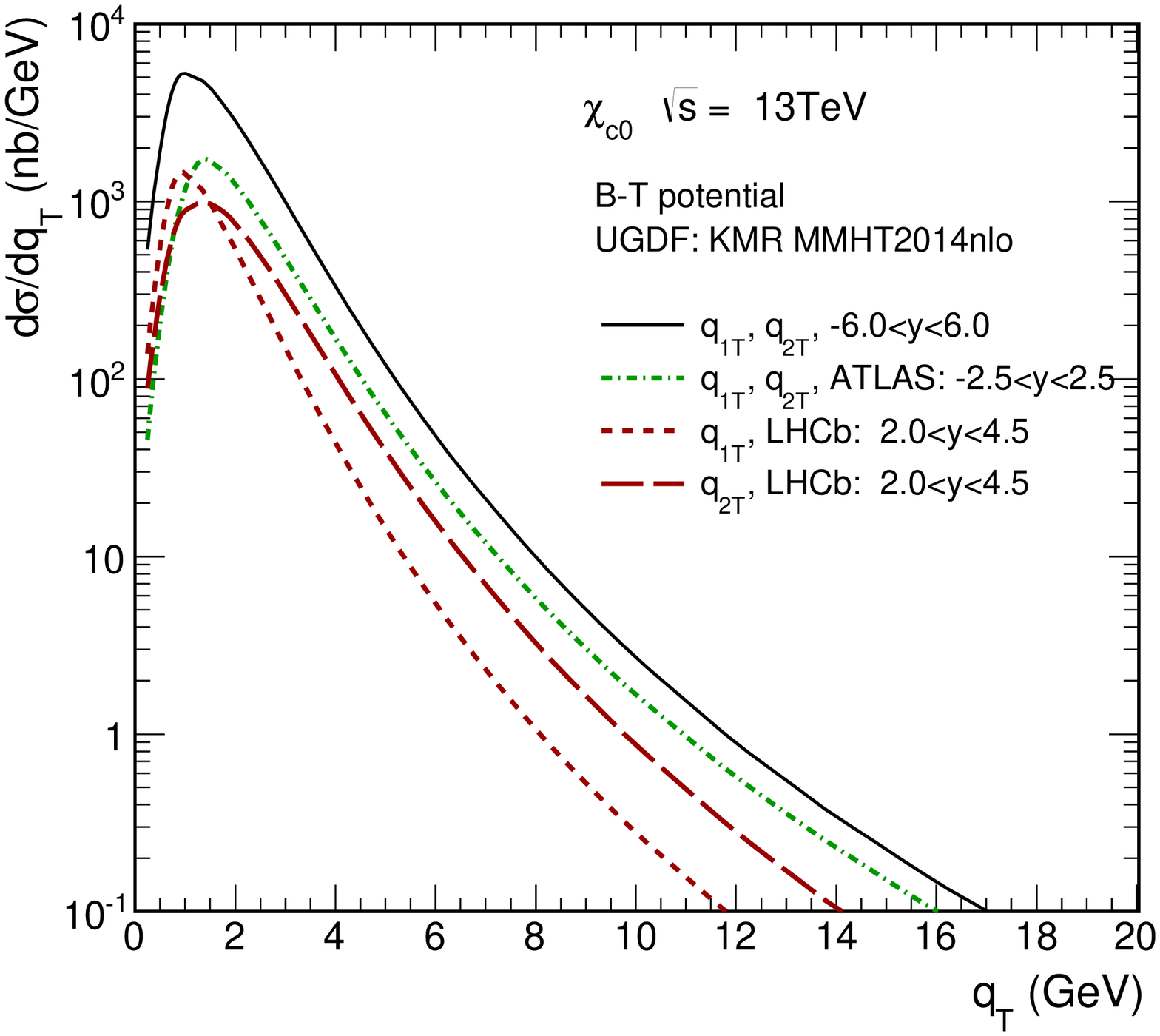}
    \includegraphics[width= 0.45\linewidth]{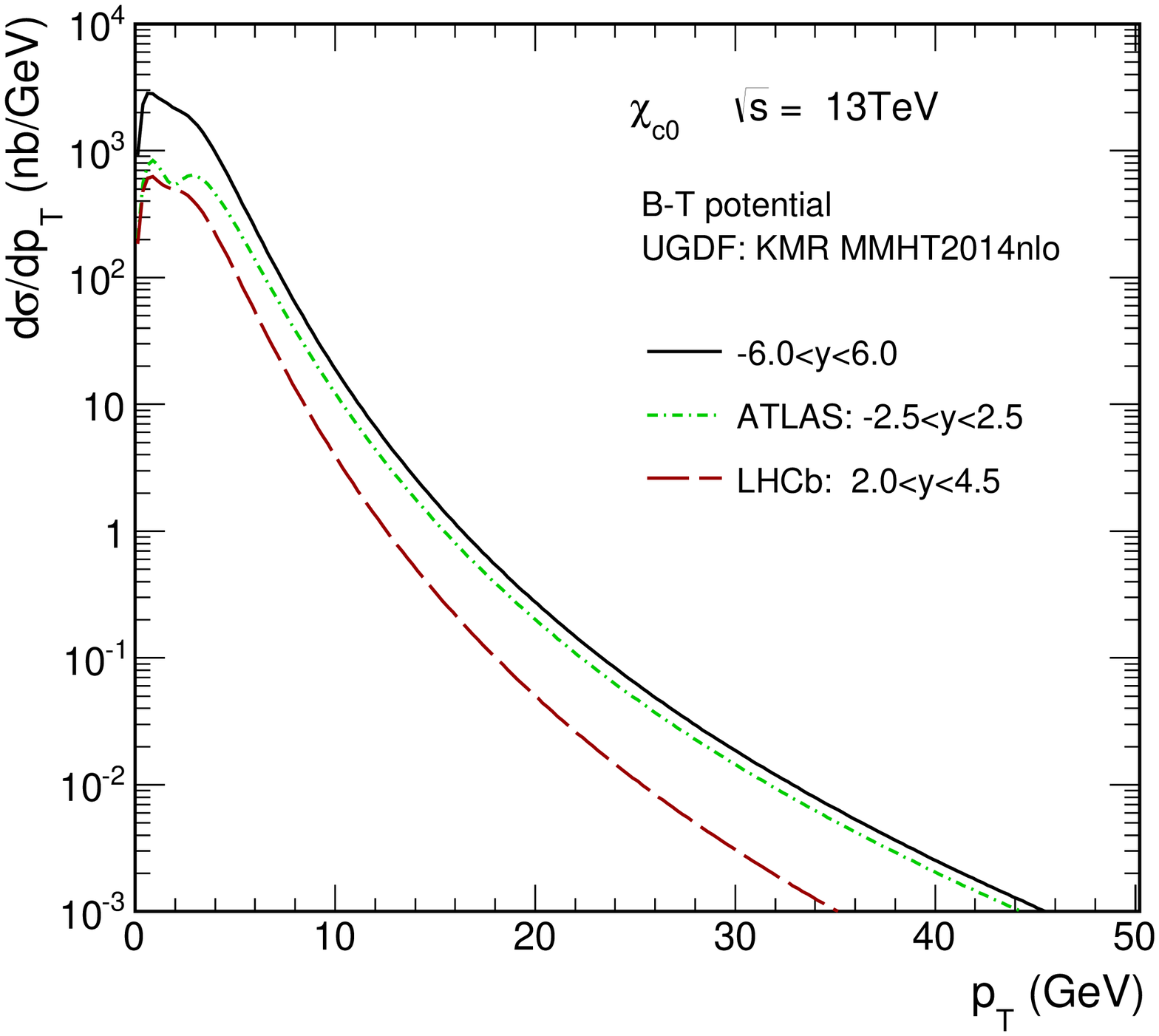}
    \caption{On the lhs differential cross section as a function of 
      the transverse momentum of initial gluon ($q_{1T}$ or $q_{2T}$) 
      with rapidity cuts for the ATLAS and LHCb experiments. 
      On the rhs differential cross section as a function of the
      transverse momentum of $\chi_{c}$ with rapidity cuts 
      for the ATLAS and LHCb experiments.}
    \label{fig:ATLAS_LHCB_chic}
\end{figure}

\begin{figure}
    \centering
    \includegraphics[width=0.45\linewidth]{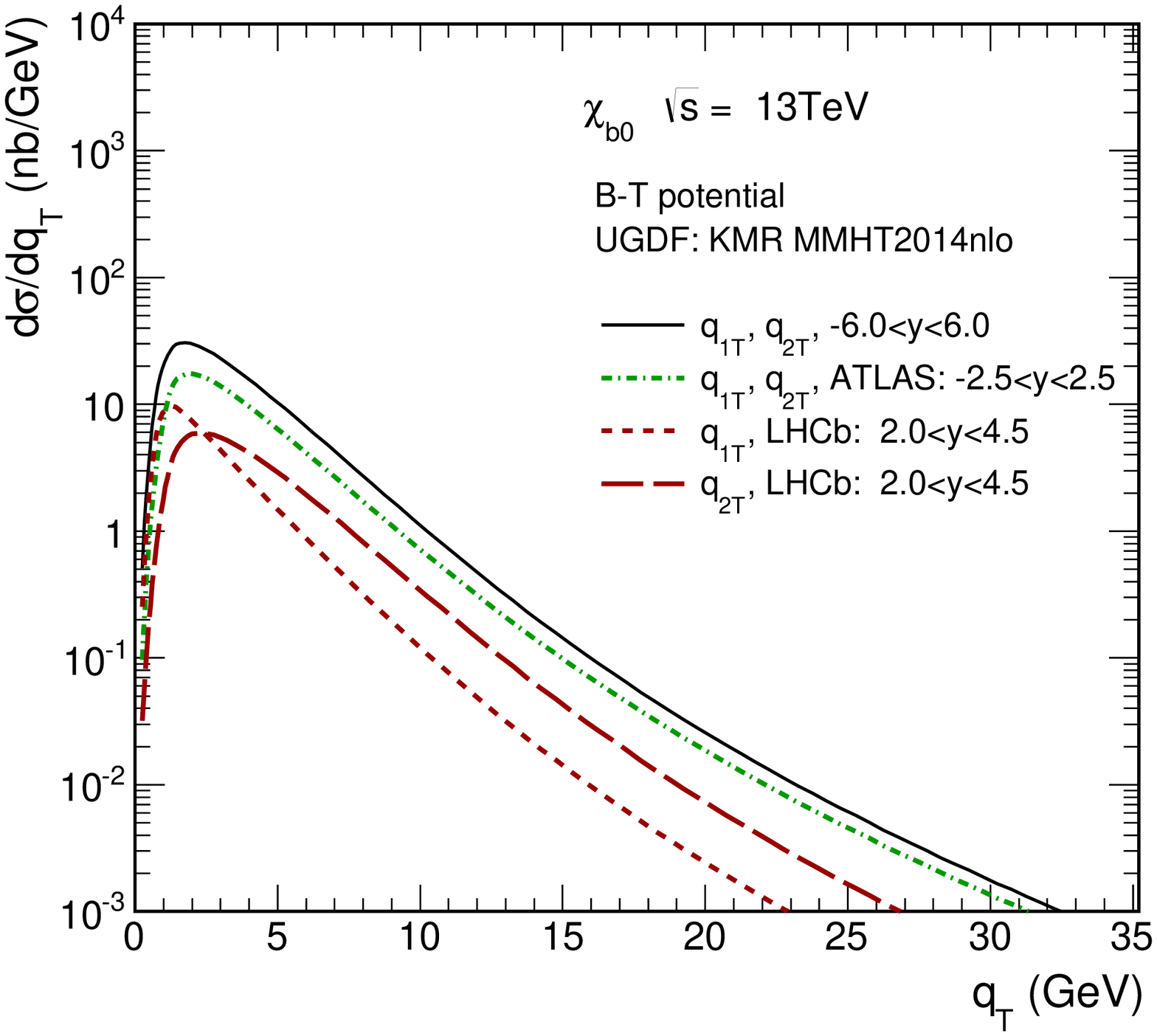}
    \includegraphics[width= 0.45\linewidth]{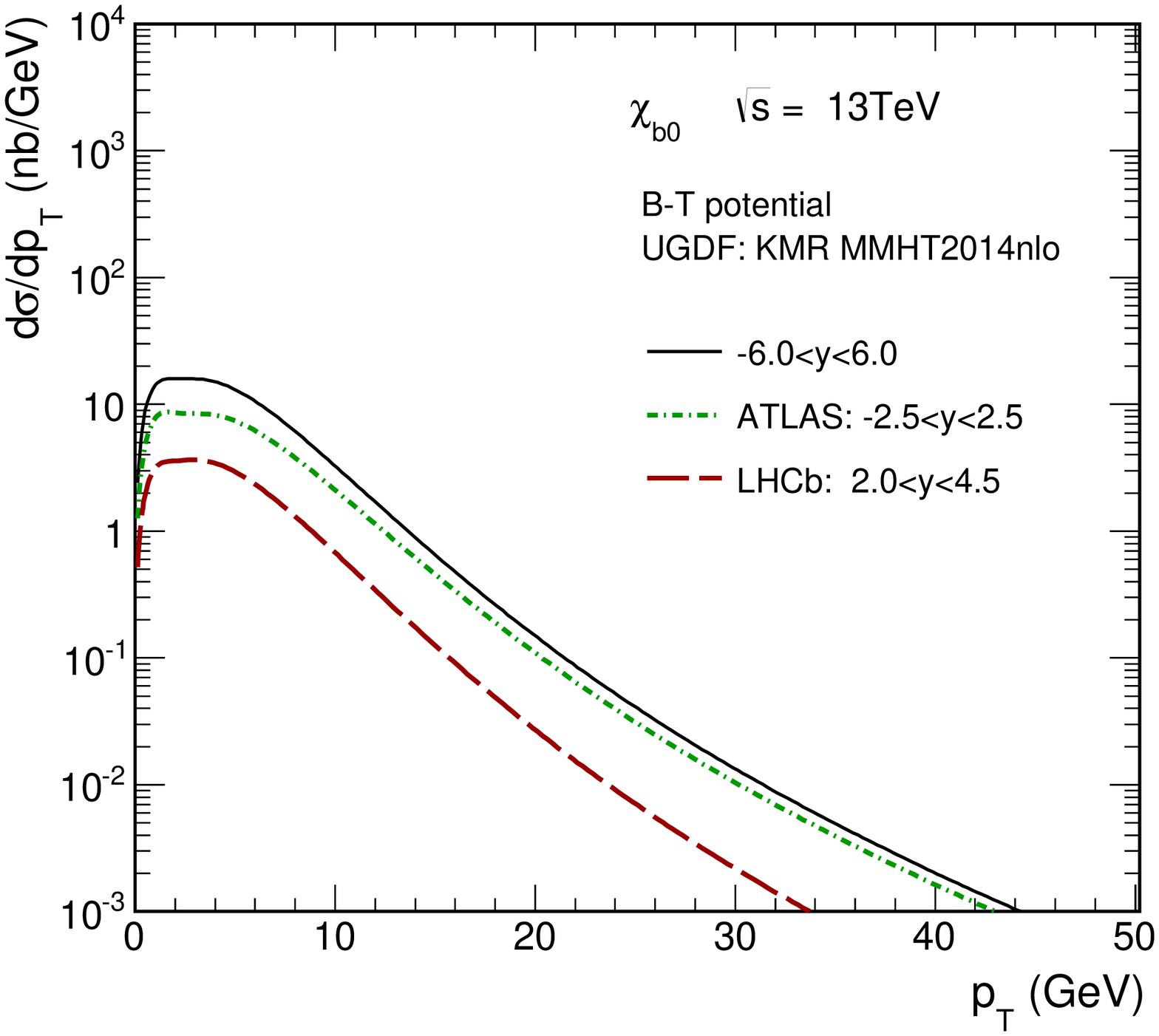}
    \caption{Differential cross section as a function of the transverse
      momentum of initial gluon ($q_{1T}$ or $q_{2T}$) with rapidty cuts 
      for the ATLAS and LHCb experiments. 
      On the rhs differential cross section as a function of the
      transverse momentum of $\chi_{b}$ with rapidity cuts 
      for the ATLAS and LHCb experiments.}
    \label{fig:ATLAS_LHCB_chib}
\end{figure}

Figs. \ref{fig:ATLAS_LHCB_chic},  \ref{fig:ATLAS_LHCB_chib} are devoted 
to specific experiments: ATLAS and LHCb. 
On the l.h.s. we show the $q_{1T}$ or $q_{2T}$ distributions.
For the ATLAS experiment they are the same, but differ for the LHCb
experiment (compare the red dashed and red dotted lines). 
On the rhs we show the $\chi_{c0}$ transverse momentum distributions
for the two experiments. A broader distribution is obtained for the
ATLAS kinematics.
The results presented on the rhs could be directly confronted with
experimental data provided such a measurement is feasible.

Let us now come to the discussion of some of the most striking results of our work,
related to the polarizations of gluons.
Indeed, Eq.(\ref{eq:gluongluon_amplitude}) to Eq.(\ref{eq:G_TT_G_LL}) clearly show the nontrivial helicity structure of the (reggeized) gluon-gluon fusion process. Differently from the collinear approach, gluons do not only carry transverse polarizations, but also longitudinal gluons enter. The $\chi$-production cross section rather encodes the properties of a whole density matrix in the space of gluon polarizations.

\begin{figure}
    \centering
    \includegraphics[width = 0.45\linewidth]{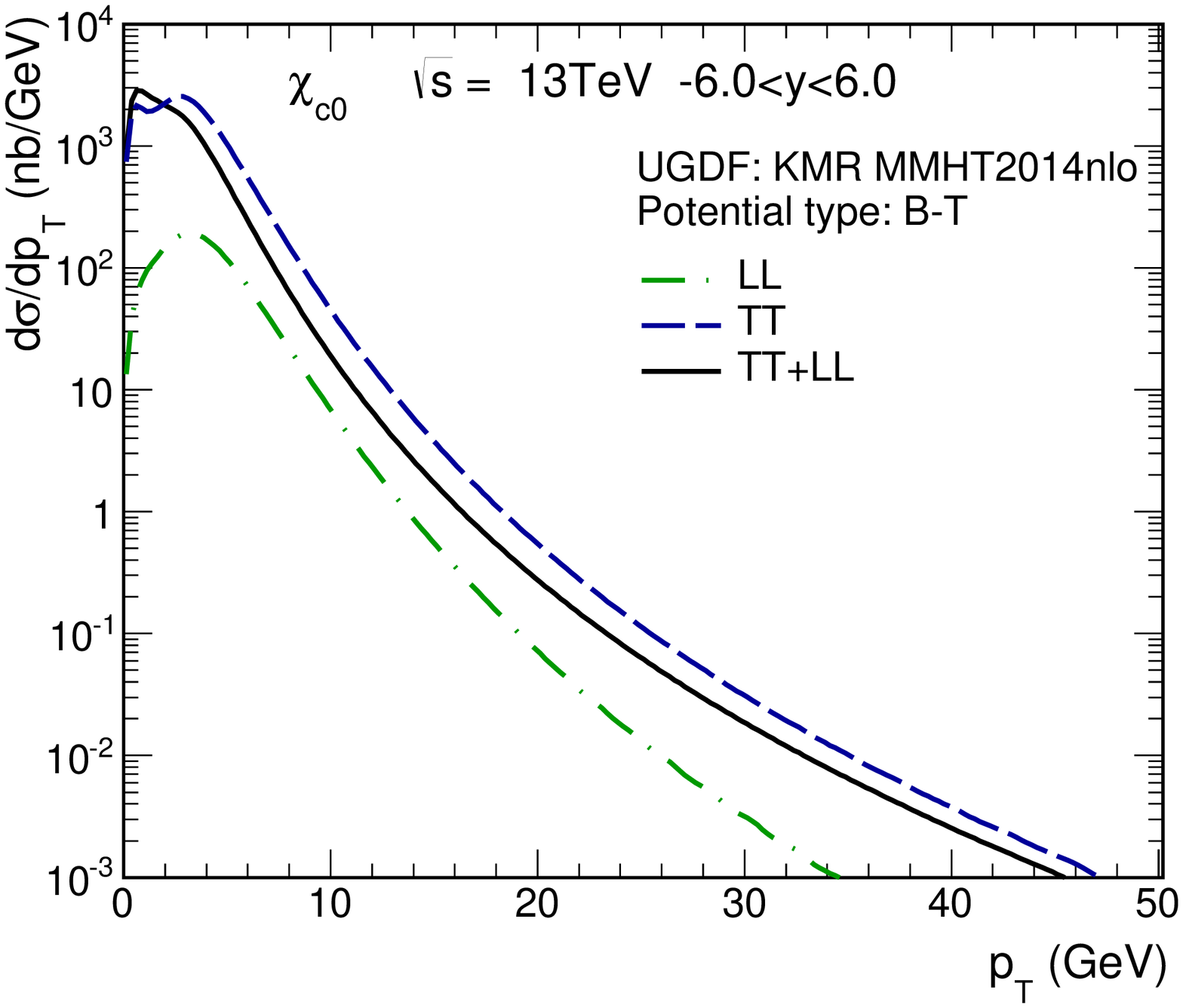}
    \includegraphics[width = 0.45\linewidth]{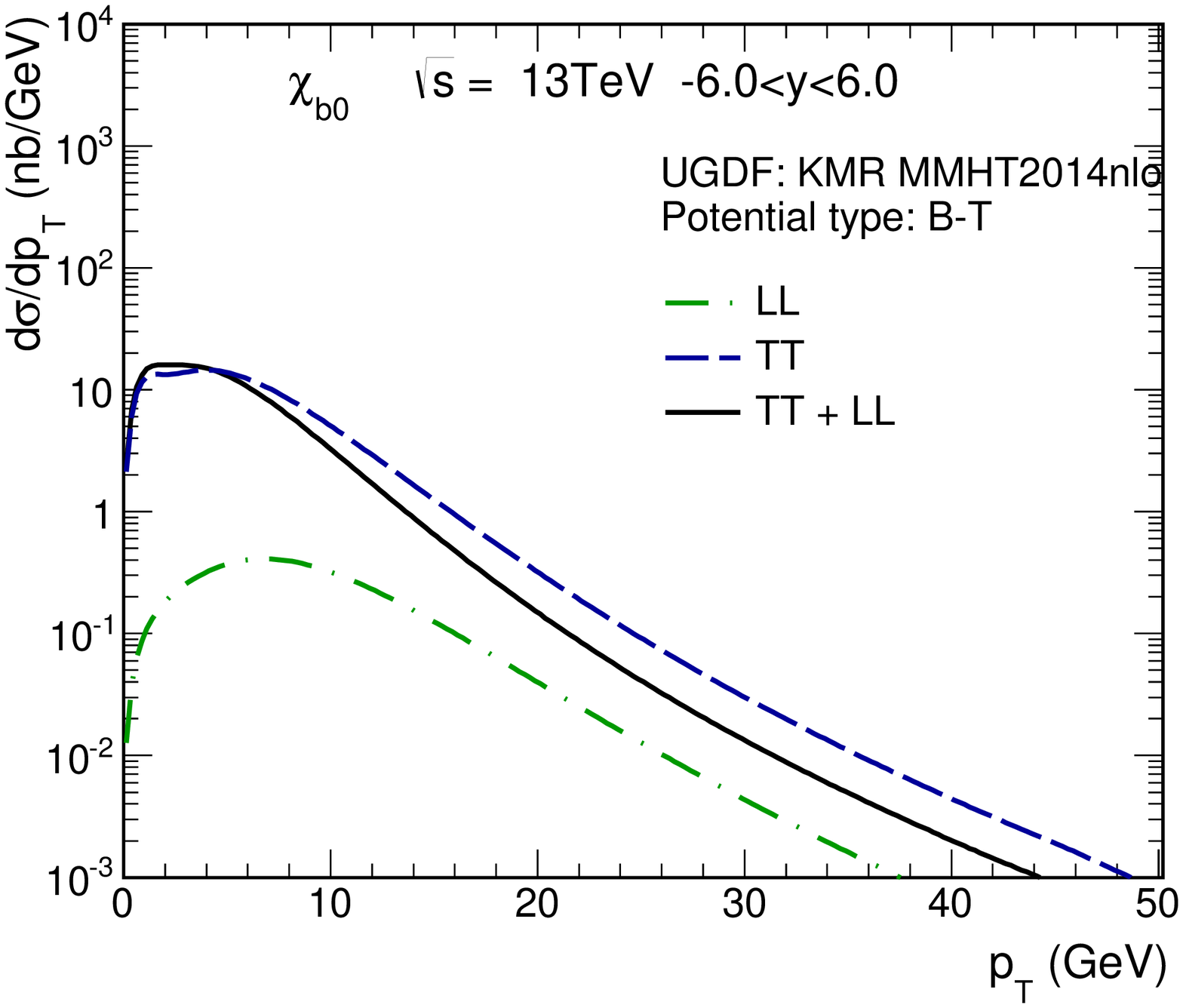}
    \caption{The comparison of meson transverse momentum distribution
    with only the $LL$ amplitude and only $TT$ amplitude as well as for the full, coherent sum of $TT$ and $LL$  amplitudes. Here the  B-T potential was used.}
    \label{fig:dsig_dpt_tau}
\end{figure}

\begin{figure}
    \centering
    \includegraphics[width = 0.45\linewidth]{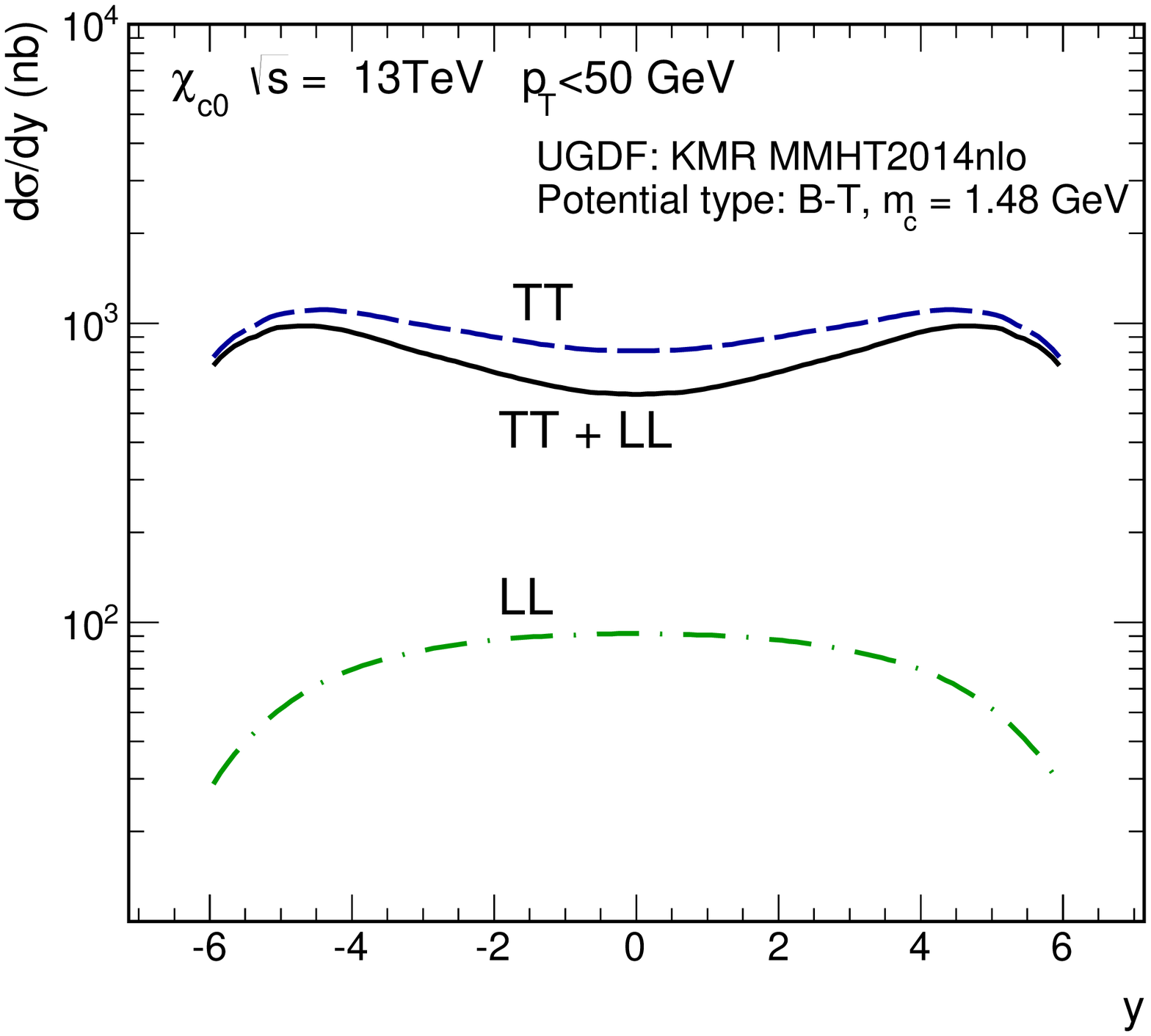}
    \includegraphics[width = 0.45\linewidth]{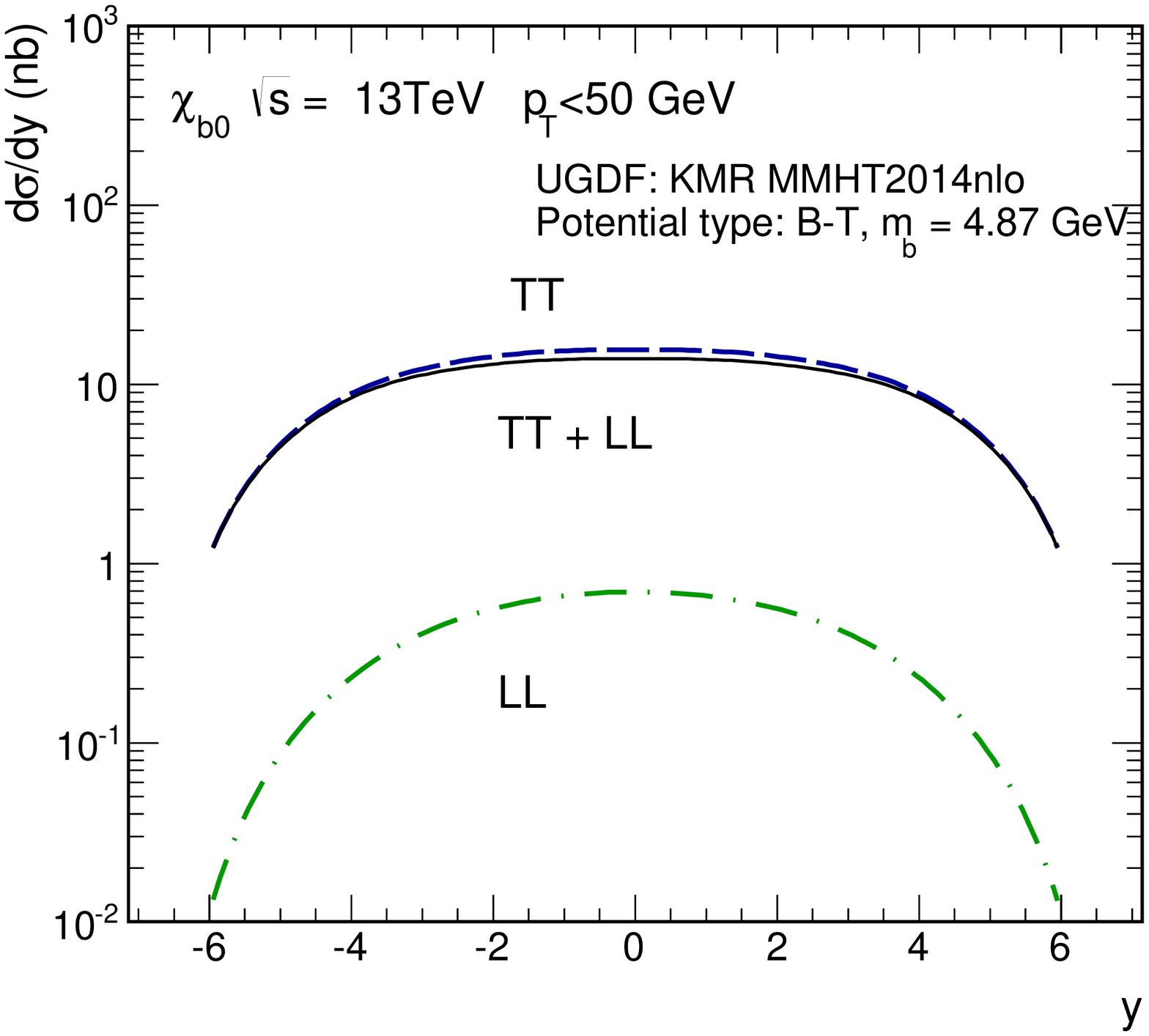}
    \caption{The comparison of meson rapidity distribution
    with only $LL$ amplitude and only $TT$ amplitude as
    well as for the full coherent sum of $TT$ and $LL$ amplitudes. Here the B-T potential was used.}
    \label{fig:dsig_dy_tau}
\end{figure}

In Fig.~\ref{fig:dsig_dpt_tau} and Fig.~\ref{fig:dsig_dy_tau} we compare contributions from TT and LL amplitudes of the process
$pp \to \chi_{c0}$ and $pp \to \chi_{b0}$ to their coherent sum.
The LL component when separate is an order of magnitude smaller than
the TT one. However, one can observe net negative interference effect -- the
cross section when including the LL component is smaller than that
obtained for the TT component only. This is clearly visible
for rapidity distributions for $\chi_{c0}$.
The details depend, however, on the meson transverse momentum.
The interference effect changes sign at small $p_T$.

\section{Conclusions}
\label{sec:Conclusions}

We have presented a formalism how to include $c \bar c$ wave functions
for calculating production of $\chi_{c0}$ and $\chi_{b0}$ P-wave
quarkonia in proton-proton collisions within the $k_T$-factorization approach.
The corresponding radiative decay widths were discussed in detail.
We have presented a discussion about the relevant quark mass dependence
of the form factors at $Q_1^2 = 0$ and $Q_2^2 = 0$ and radiative decay widths.
We have presented several tables demonstrating dependences on
approximations used.

Different representations of the $\gamma^* \gamma^* \to \chi_{Q0}$
form factors were considered.
We have presented and discussed the form factors as functions of 
($Q_1^2,Q_2^2$) and ($\bar{Q^2},\omega$),
where $\bar{Q^2} = \frac{Q_1^2 + Q_2^2}{2}$ and
$\omega = \frac{Q_1^2 - Q_2^2}{Q_1^2 + Q_2^2}$. 
In contrast to the $\eta_c$ case we observe no scaling of 
the $F_{TT}$ and $F_{LL}$ form factors in $\omega$ for $\chi_{c0}$.
We have presented also the form factors as functions of single photon virtuality.
Such form factors could be measured in future by Belle2 experiments. 

Similar from factors were calculated next for $g^* g^* \to \chi_{c0}$.
The form factors have then been used to calculate
rapidity and transverse momentum distributions of $\chi_{c0}$ at
$\sqrt{s}$ = 13 TeV within the $k_T$-factorization approach.
Two different UGDFs from the literature have been used.

We have compared the results of our full calculations with
the form factors obtained with the light-cone $c \bar c$ wave function 
to the results of the NRQCD calculations as well as with those 
for $\eta_c(1S)$ production obtained recently \cite{BPSS2019}.
The cross section for $\chi_{c0}$ production is somewhat smaller
than that for the $\eta_c(1S)$ production which makes a measurement
in the $p \bar p$ channel rather difficult. The $\gamma \gamma$ or 
$\gamma J/\psi$  final states might be a better option.

Finally we have performed a decomposition of the cross sections for
$p p \to \chi_{c0}$ and $p p \to \chi_{b0}$ into TT, LL and their
interference components. On average we have found a negative
interference effect. For $\chi_{c0}$ production the inclusion of the LL
components lowers the rapidity distributions - the result of the
complete calculation is lower than that for the TT component alone.
The interference effect depends, however, on $\chi_{c0}$ or
$\chi_{b0}$ transverse momenta.

\vspace{1cm}

{\bf Acknowledgements}

This work was partially supported by the Polish National Science Center
grant UMO-2018/31/B/ST2/03537 and by the Center for Innovation and
Transfer of Natural Sciences and Engineering Knowledge in Rzesz{\'o}w.
This work was supported by Polish National Agency for Academic Exchange
under Contract No. PPN/IWA/2018/1/00031/U/0001. R.P. is partially supported 
by the Swedish Research Council grant No. 2016-05996, 
by the European Research Council (ERC) under the European Union's Horizon 2020 
research and innovation programme (grant agreement No 668679), as well as by 
the Ministry of Education, Youth and Sports of the Czech Republic project LTT17018 
and by the NKFI grant K133046 (Hungary).
Part of this work has been performed in the framework of COST Action CA15213 “Theory
of hot matter and relativistic heavy-ion collisions” (THOR).

\appendix

\section{Light-cone wave function for 
the scalar meson $^{2S+1}L_J = ^3P_0, \, J^{PC} = 0^{++}$}
\label{sec:LCWF}

To construct the light-front wave function of the $Q \bar Q$ bound state
of good angular momentum quantum numbers we use the procedure based on the
Melosh tranformation \cite{Melosh:1974cu,Jaus:1989au} and Terentev substitution \cite{Terentev:1976jk} starting from the quark-model
rest-frame wave function (WF). 
For the scalar meson, which is a $J=0$, $L=1,S=1$ state, the rest-frame WF
has the form
\begin{eqnarray}
\Psi_{\tau \bar \tau} (\vec k) &=&   \sum_{L_z,S_z} Y_{1L_z}(\hat k) \bra{\half \half \tau \bar \tau} | \ket{1 S_z} \bra{11 L_z S_z}|\ket{0 0} {u(k) \over k}
\nonumber \\
&=& {1 \over \sqrt{2}} \xi^{\tau \dagger}_Q \, {(\vec \sigma \cdot \vec k) \over k} i \sigma_2 \, \xi^{\bar \tau *}_{\bar Q} \, {u(k) \over k} {1 \over \sqrt{4 \pi}}.
\label{eq:RF_WF}
\end{eqnarray}
Here, following the notation of \cite{Cepila:2019skb}, $\xi_Q^\tau, \xi_{\bar Q}^{\bar \tau *}$ are the canonical two-spinors of quark and antiquark respectively.
The wave function is normalized as
\begin{eqnarray}
 \int d^3 \vec k \, \sum_{\tau \bar \tau} | \Psi_{\tau \bar \tau} (\vec k) |^2 = 1,
\end{eqnarray}
which implies for the radial WF $u_1(k)$, that
\begin{eqnarray}
 \int_0^\infty dk  \, u^2(k) = 1. 
\end{eqnarray}
We now wish to express the spin-orbit structure in terms of light front spinors
$\chi_Q^\lambda, \chi_{\bar Q}^{\bar \lambda *}$, which are related to the 
canonical spinors via the Melosh-transform:
\begin{eqnarray}
 \xi_Q &=& R(z,\bk) \chi_Q \nonumber \\
 \xi^*_{\bar Q} &=& R^*(1-z,-\bk) \chi^*_{\bar Q} \, ,
\end{eqnarray}
with the unitary matrix $R(z,\bk)$
\begin{eqnarray}
R(z, \bk) = { m_Q + z M - i \vec \sigma \cdot ( \vec n \times \bk) \over \sqrt{ (m_Q + z M)^2 + \bk^2  }} = { m_Q + z M - i \vec \sigma \cdot ( \vec n \times \bk) \over \sqrt{ zM (M+2m_Q) }} \, .
\end{eqnarray}

Here $z$ and $1-z$ are the fractions of the meson's lightfront plus-momentum  
carried by the quark and antiquark respectively, while their transverse momenta
are denoted by $\pm \bk$.
We also introduced $\vec{n} = (0,0,1)$ and the vector product $\vec n \times \bk = (-k_y,k_x,0)$.
The three-momentum $\vec k$ in the $Q \bar Q$  rest frame is parametrized as
\begin{eqnarray}
 \vec k = (\bk, k_z) = (\bk, \half (2z-1) M) \, ,
\end{eqnarray}
where $M$ is the invariant mass of the $Q \bar Q$ system obtained from
\begin{eqnarray}
 M^2 = {\bk^2 + m_Q^2 \over z (1-z)}.
\end{eqnarray}
Notice also, that
\begin{eqnarray}
 k = |\vec k| = \half \sqrt{M^2 - 4m_Q^2} \, .
\end{eqnarray}
The light-front wave function depending on light-front helicities $\lambda \bar \lambda$ now has the form
\begin{eqnarray}
 \Psi_{\lambda \bar \lambda} (z,\bk) = \chi_Q^{\lambda \dagger}
 \, {\cal O}' \, i \sigma_2 \, \chi_{\bar Q}^{\bar \lambda *}  \,
 \phi(z, \bk) \; .
\label{eq:LFWF}
\end{eqnarray}
Here the radial light-front wave function is expressed as
\begin{eqnarray}
 \phi(z,\bk) = {u(k) \over k} \, {\sqrt{{\cal J}} \over
   \sqrt{4 \pi}} \; ,
\end{eqnarray}
where ${\cal J}$ is a jacobian
for the transition from integration over $\vec k$ to the relativistic two-body phase space parametrized through $z, \bk$:
\begin{eqnarray}
 d^3 \vec p = {\cal J} {dz d^2 \bk \over z(1-z) 16 \pi^3} \, ,
\end{eqnarray}
which gives $\sqrt{{\cal J}} = 2 \sqrt{M \pi^3}$, so that 
\begin{eqnarray}
 \phi(z,\bk) = \pi \sqrt{M} {u(k) \over k}.
\end{eqnarray}
Let us turn now to the spin-orbit factor. 
Denoting ${\cal O} = \vec \sigma \cdot \vec k/(\sqrt{2}k)$, we can write the Melosh-transformed spin-orbit vertex as
\begin{eqnarray}
 {\cal O}' = R^\dagger(z,\bk) {\cal O} \, i \sigma_2
 R^*(1-z,-\bk) (i \sigma_2)^{-1} =  
R^\dagger(z,\bk) {\cal O} \, R(1-z,-\bk) \, .
\end{eqnarray}
In the last equality, we used the property of Pauli-matrices
\begin{eqnarray}
 i \sigma_2  \, \vec \sigma^* \, (i \sigma_2)^{-1} = - \vec \sigma \, .
\end{eqnarray}
Inserting the explicit form of matrices $R$, we obtain
\begin{eqnarray}
 {\cal O}' &=& {1 \over \sqrt{z (1-z)}} {1 \over M (M+ 2m_Q) } 
 \Big\{ (m_Q^2 + m_Q M + z(1-z)M^2) {\cal O} \nonumber\\
 && - \vec \sigma \cdot ( \vec n \times \bk) \, {\cal O} \, \vec \sigma \cdot ( \vec n \times \bk) \nonumber \\
 && +  (M+2m_Q) { i \over 2} \Big( {\cal O} \, \vec \sigma \cdot ( \vec n \times \bk) + \vec \sigma \cdot ( \vec n \times \bk) \, {\cal O}\Big) 
 \nonumber \\
 &&+ (2z-1) M {i \over 2}  \Big( {\cal O} \, \vec \sigma \cdot ( \vec n \times \bk) - \vec \sigma \cdot ( \vec n \times \bk) \, {\cal O}\Big) \Big\}.
\end{eqnarray}
For the case of interest, using the well known properties of Pauli matrices, we 
arrive at
\begin{eqnarray}
 {\cal O}' = {1 \over \sqrt{z(1-z)}} {1 \over 2 \sqrt{2} k} \Big\{ \vec
 \sigma \cdot \bk + (2z-1) m_Q \, \vec \sigma \cdot \vec n \Big\} \, .
\end{eqnarray}
We note in passing, that the effect of the Melosh rotation is simply a substitution of
\begin{eqnarray}
 \vec k \to \vec k' = \Big( {\bk \over 2 \sqrt{z(1-z)}}, {(2z-1) m_Q \over 2 \sqrt{z(1-z)}}  \Big)\, , 
\end{eqnarray}
in the spin-orbit part of Eq.(\ref{eq:RF_WF}).

Inserting this into Eq.(\ref{eq:LFWF}) gives us the following components
of the light-front wave function:
	\begin{eqnarray}
	\Psi_{\lambda \bar \lambda}(z,\bk) &=&  \begin{pmatrix}
	\Psi_{++}(z,\bk) & \Psi_{+-}(z,\bk) \\
	\Psi_{-+}(z,\bk) & \Psi_{--}(z,\bk)
	\end{pmatrix} \nonumber \\
	&=& {-1 \over \sqrt{z(1-z)}} 
	\begin{pmatrix}
	k_x - i k_y & m_Q(1-2z) \\
	m_Q(1-2z) & -k_x-ik_y
	\end{pmatrix} 
	{\phi(z,\bk) \over \sqrt{2}\sqrt{M^2 -4m_Q^2}} \, .
	\label{eq:Psi_matrix}
	\end{eqnarray}
To lighten up the notation in the rest of the paper, we further absorb a factor 
$1/(2 \sqrt{2} k)$ into the wave function, and introduce
\begin{eqnarray}
 \psi(z,\bk) \equiv {\phi(z,\bk) \over
   \sqrt{2}\sqrt{M^2 -4m_Q^2}} \, .
\end{eqnarray}
The normalization condition reads
\begin{eqnarray}
 1 &=& \int {dz d^2 \bk \over z(1-z) 16 \pi^3} \sum_{\lambda, \bar \lambda} | \Psi_{\lambda \bar \lambda}(z, \bk) |^2 = 
 \int {dz d^2 \bk \over z(1-z) 16 \pi^3} \, |\phi(z,\bk)|^2 \nonumber \\
 &=& \int {dz d^2 \bk \over z(1-z) 16 \pi^3} \, 2 (M^2 - 4m_Q^2) \, |\psi(z,\bk)|^2 \, .
\end{eqnarray}


\end{document}